\documentclass[aps,preprint,superscriptaddress]{revtex4-1}
\usepackage{graphicx}
\usepackage[english]{babel}
\usepackage{booktabs}
\usepackage{amsmath}
\usepackage{latexsym}
\usepackage{amssymb}
\usepackage{amscd}
\usepackage{array}
\usepackage{tabularx}
\usepackage{longtable}
\usepackage{subfigure}

\allowdisplaybreaks[1]

\def\e{{\rm e}}
\def\g{\boldsymbol}

\def\ssc{\scriptscriptstyle}
\def\scr{\scriptstyle}

\begin{document}
\title{Description of cross peaks induced by intermolecular vibrational energy transfer in 
two-dimensional infrared spectroscopy}
\author{Albert A.\ Villaeys}
\affiliation{Research Center for Applied Sciences, Academia Sinica, Taipei 115, Taiwan}
\affiliation{Universit\'e de Strasbourg et Institut de Physique et Chimie des Mat\'eriaux de Strasbourg, France}
\author{Kuo Kan Liang}\email{kkliang@sinica.edu.tw}
\affiliation{Research Center for Applied Sciences, Academia Sinica, Taipei 115, Taiwan}
\affiliation{Department of Biochemical Science and Technology, National Taiwan University, Taipei 106, Taiwan}

\begin{abstract}

In the present work, the analytical description of an intermolecular vibrational energy transfer, analyzed by two dimensional infrared spectroscopy, is established. The energy transfer process takes place between the dark combination states of low frequency modes pertaining to different molecules. The appearance of the cross peaks results from coherent transfer between these combination states and an optically active state of the acceptor molecule. Such a process has recently been observed experimentally between the nitrile groups of acetonitrile-$_{\rm d3}$ and  benzonitrile molecules. This molecular system will be used as a model for the simulations of their two-dimensional infrared spectra. The dependence of the cross-peak growth, which is a signature of the intermolecular energy transfer, will be discussed in detail as a function of the molecular dynamical constants.  
\end{abstract}
\maketitle 

\section{ Introduction}

Intermolecular vibrational energy transfer occurs in a quite large number of chemical processes ranging from simple chemical reactions to very complicated molecular systems like molecular motors. It has been the subject of numerous publications \cite{bb01,bb02,bb03,bb04,bb05,bb06}.  
Among the large variety of experimental techniques used to investigate intramolecular or intermolecular energy transfer, two important spectroscopic methods have been quite successful these last decades. The first one is based on the infrared pump-probe spectroscopy technique. It enables an indirect measurement of the intermolecular energy transfer by bleaching the population at the pumping laser frequency and testing the subsequent ground state recovery by a time delayed probe pulse. A typical situation is the one reported by Bakker {\it et al} \cite{bb07}
on the study of vibrational energy transfer, after excitation of the C-H stretching mode in chloroform and bromoform dissolved in polar and non-polar solvent, where an important increase of vibrational energy transfer resulting from the molecule-solvent dipole-dipole interaction has been observed. Similar conclusions have been reached for resonant intermolecular vibrational energy transfer in liquid water from rotational anisotropy measurements \cite{bb08}.
The second archetype of experiments takes advantage of the proportionality between the time-dependence of the Raman active transition and the population time dependence to apply infrared pump coupled to anti-Stokes-Raman-probe technique to study vibrational energy redistribution in polyatomic liquids such as water, deuterated water and methanol \cite{bb09}.
 
Today, ultrafast two-dimensional infrared (2D-IR) spectroscopy which has been used for more than a decade to study dynamics and structures of large molecular complexes 
\cite{bb10,bb11,bb12}, charge transfer \cite{bb13,bb14}, chemical reactions \cite{bb15,bb16,bb17}
as well as vibrational coupling and vibrational energy relaxation \cite{bb18,bb19,bb20},
offers an interesting method to directly probe the intermolecular vibrational energy transfer \cite{bb21,bb22,bb23,bb24,bb25}. 
In addition, it does not require the probe vibrational mode to be involved in the intermolecular vibrational energy transfer and can display interactions and dynamics that are not accessible by traditional linear infrared vibrational absorption measurements. 

In a recent experimental work, Bian {\it et al} \cite{bb21}
have used 2D-IR spectroscopy to directly probe the intermolecular vibrational energy transfer between benzonitrile and acetonitrile-$_{\rm d3}$ molecules. Using an experimental setup based on three transform-limited IR femtosecond laser pulses and heterodyne detection to phase-resolve and to amplify the vibrational echo signal, they directly probe the vibrational energy transfer between the two molecules. In these 2D-IR spectroscopy experiments, three different types of variables are involved. The first one is the emitted vibrational echo frequency. The second one is the delay time between two laser pulses, most frequently the first and the second ones. It will be numerically Fourier transformed for each vibrational echo frequency to give a second frequency mapping. Finally, the third one corresponds to the time of delay between the two last laser pulses. It is just due to this last parameter that the cross peak, which is the signature of the intermolecular energy transfer, appear in 2D-IR spectra. 

When the molecular vibrational states participating in the transfer of energy are dark states, most probably made of combination states of low frequency modes with very weak dipole moments or 
not at all optically active, coherence transfer between these combination states and an optically active vibrational state induces a cross-peak which becomes the signature of the energy transfer process. It is the intention of this work to describe the appearance of this cross-peak and to analyze in detail the influence of the dynamical constants on its magnitude. In Sec.2, we introduce the vibrational structure of the individual molecules participating in the energy transfer. Then, the energy level structure of the global system undergoing the dynamics induced by the femtosecond laser pulses is built in terms of the physical constants of the individual molecules. In Sec.3, we describe the general dynamics induced by the three femtosecond laser pulses and resulting from the relaxation, dephasing, and coherence transfer processes. Then, the 2D-IR spectra will be evaluated and discussed according to the roles of the physical constants. Finally, in the last section, numerical simulations exhibit the main features which characterize the identification of the energy transfer process through the appearance of the cross-peak observed experimentally. 

\section{ General dynamics underlying a 2D-IR spectrum} 

The physical process underlying the 2D-IR vibrational spectrum of a molecular system is based on the interaction of three delayed infrared pulses with the molecular system under investigation.
As usual, the interaction Hamiltonian of the laser-system interaction is given by
\begin{eqnarray}
\g{V}(t)=-\sum_{p=a,b,c} \mathcal{A}_p(t-T_p)\bigl[\vec{\g{\mu}}\cdot\vec{\mathcal{E}}_p\e^{-i\omega_p(t-T_p)+i\vec{k}_p\cdot\vec{r}}+ {\rm C.C.}\bigr]
\label{2.1}
\end{eqnarray}    
where the notation C.C. stands for the complex conjugate part. The symbol $\mathcal{A}_p(t-T_p)$ stands for the normalized envelops of the three laser pulses defined as $\mathcal{A}_p(t-T_p)=\sqrt{\gamma_p}\exp(-\gamma_p\vert t-T_p\vert)$ with $p=a,\,b\,{\rm or\,c}$. This is a convenient form for later analytical evaluation and its time-integrated form is physically acceptable and has been used extensively. All the laser pulses are centered at $\omega_p=2230\,{\rm cm}^{-1}$ with a time duration of $\gamma_p^{-1}=55\,{\rm fs}$ and with their amplitudes arbitrarily fixed. As usual, $\vec{\g{\mu}}$ is the dipole moment operator and the quantities $\omega_p$ and  $\vec{k}_p$ are standard notations for the frequency and wave vector of the field $p$, respectively. It is worthy of note that in time-domain nonlinear spectroscopy, the pulse delay time $T_p$ are retained in the pulse envelops only, because they act just like arbitrary factors $\exp(i\omega_pT_p)$ and can be compensated by the phase of the local field oscillator for heterodyne detected signals or just cancel in homodyne detection. To analyze 2D-spectra, it is necessary to perform a Fourier transform over pulse delay time. The above mentioned factors have to be retained. Moreover, for 2D-IR experiments, it is convenient to use the photon-echo geometry. Therefore, it is also adopted in our theoretical calculation. It is well established that photon-echo and many other related nonlinear optical processes \cite{bb26,bb27,bb28,bb29,bb30,bb31,bb32,bb33}
are described by the third-order perturbation term of the density matrix with respect to the laser-molecule interaction $\g{V}(t)$ defined previously by Eq.(\ref{2.1}). Then, the contribution to the third-order term of the density matrix, $\g{\rho}^{(3)}(t)$, that is relevant to the three-pulse process discussed here, takes the form
\begin{multline}
\g{\rho}^{(3)}(t)=\frac{i}{\hbar^3}\int_{t_0}^{t}\!\!\!d\tau_3\int_{t_0}^{\tau_3}\!\!\!d\tau_2\int_{t_0}^{\tau_2}\!\!\!d\tau_1\g{G}(t-\tau_3)\g{L}_v(\tau_3)\g{G}(\tau_3-\tau_2)\g{L}_v(\tau_2)\g{G}(\tau_2-\tau_1) \g{L}_v(\tau_1)\g{\rho}(t_0)\\
\label{2.2}
\end{multline}
where the interaction Liouvillian is defined by $\g{L}_v(\tau_i)=[\g{V}(\tau_i),\cdots]$. The various $\g{G}(\tau_i-\tau_j)$ are free evolution Liouvillians of the vibrational system, corresponding to $\g{G}(\tau_i-\tau_j)=\e^{-\frac{i}{\hbar}\g{L}(\tau_i-\tau_j)}$ if $\g{L}=[\g{H},\cdots]$ with $\g{H}$ the free vibrational model Hamiltonian. 
The evolution Liouvillian $\g{G}_{mmnn}(\tau_i-\tau_j)$ account for the evolution of the populations from $n\rightarrow m$. The other two evolution Liouvillians, namely $\g{G}_{mnmn}(\tau_i-\tau_j)$ with $m\not=n$ and $\g{G}_{mrms}(\tau_i-\tau_j)$ with $m\not=r,s$ and $r\not=s$ account for the evolution of the coherences. Notice that the last Liouvillian describes coherence transfer between the quasi-resonant transitions $m\rightarrow r$ and $m\rightarrow s$. The emitted signal along the rephasing phase-matched direction $\vec{k}_{reph}=-\vec{k}_{a}+\vec{k}_{b}+\vec{k}_{c}$ and the nonrephasing direction $\vec{k}_{nonreph}=\vec{k}_{a}-\vec{k}_{b}+\vec{k}_{c}$ is deduced from the polarization 
\begin{eqnarray}
\vec{P}^{(3)}_{\vec{k}_s}(T_a,T_b,T_c,t)=2\Re\Big[\sum_{i}\sum_{j<i}\g{\rho}^{(3)}_{\vec{k}_s,ij}(t)\vec{\g{\mu}}_{ji}\Big]  
\label{2.3}
\end{eqnarray}
where $\vec{k}_s$ stands for $\vec{k}_{reph}$ or $\vec{k}_{nonreph}$. Heterodyne detection of nonlinear signals has been discussed extensively by Joffre {\it et al} \cite{bb34,bb35,bb36}.    
and will be introduced to amplify the emitted signal. To this end, an appropriate local field oscillator is introduced
\begin{eqnarray}
\vec{E}_{lo}(t)=\mathcal{A}_{lo}(t-T_{lo})\bigl[\vec{\mathcal{E}}_{lo}\e^{-i\omega_{lo}(t-T_{lo})+i\vec{k}_{lo}\cdot\vec{r}-i\Psi}+ {\rm C.C.}\bigr]
\label{2.4}
\end{eqnarray}
where the various constants are similar to the ones introduced for the exciting laser fields and $\Psi$ is an additional phase of this local field. Also, we use the relation between polarization $\vec{P}^{(3)}_{\vec{k}_s}(T_a,T_b,T_c,t)$ and signal field $\vec{E}_{\vec{k}_s}(T_a,T_b,T_c,t)$ obtained from Maxwell equations \cite{bb37,bb38},
say
\begin{eqnarray}
\vec{E}_{\vec{k}_s}(T_a,T_b,T_c,t)\propto i \vec{P}^{(3)}_{\vec{k}_s}(T_a,T_b,T_c,t).
\label{2.5}
\end{eqnarray}
To simulate the experimental procedure where the Fourier transform and next the squared magnitude are performed successively, the detected heterodyne signal intensity can be expressed as \cite{bb39}  
\begin{eqnarray}
I_{\vec{k}_s}(T_a,T_b,T_c)=\bigg\vert \int_{-\infty}^{+\infty} dt \big\lbrack \vec{E}_{lo}(t-T_{lo})+ \vec{E}_{\vec{k}_s}(T_a,T_b,T_c,t)\big\rbrack \e^{i\omega_t t}\bigg\vert^2  
\label{2.6}
\end{eqnarray}
using heterodyne detection. If, as usual, the intensity of the signal field is negligible and the one of the local field is subtracted from the total intensity, then 
\begin{eqnarray}
I_{\vec{k}_s}(T_a,T_b,T_c,\omega_t)=2\Re\Bigg\lbrack \bigg(\int_{-\infty}^{+\infty} dt \vec{E}_{lo}(t-T_{lo})\e^{i\omega_t t}\bigg)^{\star}\bigg(\int_{-\infty}^{+\infty} dt \vec{E}_{\vec{k}_s}(T_a,T_b,T_c,t)\e^{i\omega_t t}\bigg) \Bigg\rbrack
\label{2.7}
\end{eqnarray}
Three conditions were imposed according to the assumptions made by Lepetit {\it et al} \cite{bb40}.
First, the laser pulses present clean fast leading edges. Second, time origin can be chosen for convenience, and here it is fixed at the center of the last pulse. Third, the signal field obeys the causality principle. These conditions imply that $\vec{E}_{\vec{k}_s}(T_a,T_b,T_c,t)$ is negligible for $t<0$ so that the Fourier transform with respect to time can be performed over positive time only. Notice that, all high frequency terms are neglected.

For our purpose, the evaluation of the 2D-IR spectrum requires the calculation of the double Fourier transform over the delay time $\tau=T_b-T_a$ between the pulses and the experimental time $t$. In the first Fourier transform over $\tau$, the delay time $\tau$ changes sign. Therefore, the experimental waiting time $T$ has to be defined as $T={\rm min}(\vert T_a \vert,\vert T_b \vert)$, according to the prescription of Jonas \cite{bb41}.
Since the center of the third laser pulse is fixed at the time origin, the 2D-IR spectra can be obtained as
\begin{eqnarray}
I_{\vec{k}_s}(\omega_d,\omega_t,T)==\int_{-\infty}^{\infty} \e^{-i\omega_d\tau}I_{\vec{k}_s}(\tau,T,\omega_t)      .  
\label{2.8}
\end{eqnarray}
and this is the basic quantity for evaluating any 2D-IR spectra. 

The formal expression of the third-order polarization in the phase-matched directions $\vec{k}_s=\vec{k}_{reph}$ or $\vec{k}_{nonreph}$, required for simulating any 2D-IR experiment, is given in the basis set $\{\vert j\rangle\}$ of the total vibrational molecular system. From Eq.(\ref{2.2}), the density matrix elements can be written as
\begin{eqnarray}
\g{\rho}_{ij}^{(3)}(t)&=&\frac{i}{\hbar^3}\sum_{\{n\}}^{'}\sum_{r,q,p}\int_{t_0}^{t}d\tau_3\int_{t_0}^{\tau_3}d\tau_2\int_{t_0}^{\tau_2}d\tau_1   \nonumber\\
\hskip 2truecm&&\times\mathcal{A}_r(\tau_1-T_r)\mathcal{A}_q(\tau_2-T_q)\mathcal{A}_p(\tau_3-T_p)\g{R}_{n,ij}(\tau_1,\tau_2,\tau_3,t)
\e^{-i\vec{k}_s\cdot\vec{r}}.
\label{2.9}
\end{eqnarray}
Each $n_p$ specifies a particular pathway in the Liouvillian space of the total molecular system contributing to the density matrix elements $\g{\rho}_{ij}^{(3)}(t)$.
The symbol $\sum_{r,q,p}$  stands for the summation over the various combinations of fields. These combinations must satisfy the phase-matching condition in the framework of the rotating wave approximation. That is, only the combinations of field components satisfying the secular approximation are retained. In addition, the symbol $\sum_{\{n_p\}}^{'}$ means that only the values of $n_p$ associated with the particular density matrix element $\g{\rho}_{ij}^{(3)}(t)$ must be included. At least for markovian systems, the general mathematical structure of $\g{R}_{n_p,ij} (\tau_1, \tau_2,\tau_3,t)$ is of the type
\begin{eqnarray}
\g{R}_{n_p,ij}(\tau_1,\tau_2,\tau_3,t)= Q_{n_p,r,q,p}\e^{A_{n_p,r,q,p}\tau_3+B_{n_p,r,q}\tau_2+C_{n_p,r}\tau_1}\e^{K_{n_p,r,q,p}t}
\label{2.10}
\end{eqnarray}
which enables a great simplification for further time integrations. The amplitudes $Q_{n_p,r,q,p}$ and the exponential arguments $A_{n_p,r,q,p}$, $B_{n_p,r,q}$, $C_{n_p,r}$, and $K_{n_p,r,q,p}$ will all be obtained by identification of Eq.(\ref{2.9}) with all the compatible combinations of the product $\g{G}_{ijmn}(t-\tau_3)\g{L}_{v,mnpq}(\tau_3)\g{G}_{pqrs}(\tau_3-\tau_2)\g{L}_{v,rstu}(\tau_2)\g{G}_{tutu}(\tau_2-\tau_1) \g{L}_{v,tugg}(\tau_1)\g{\rho}_{gg}(t_0)$ of matrix elements deduced from Eq.(\ref{2.2}). Each combination with specified indices $\{m,n,p,q,r,s,t,u\}$ is associated with a specific pathway $n_p$. This implies that each particular value of $n_p$ stands for one particular combination of $\{m,n,p,q,r,s,t,u\}$. All of the pathways will be evaluated in the next section according to our specific model. 

At this stage, the different matrix elements of Liouvillian evolution operator participating in free evolution have to be evaluated. Notice that the matrix elements of the free evolution Liouvillian of the type $\g{G}_{mnmn}(\tau_i-\tau_j)$ are diagonal in the coherence Liouvillian subspace and have the general form
\begin{eqnarray}
 \g{G}_{mnmn}(\tau_i-\tau_j)=\e^{-i\omega_{mn}(\tau_i-\tau_j) -\g{\Gamma}_{mnmn}(\tau_i-\tau_j)} 
\label{2.11}
\end{eqnarray}
where, as usual, $\omega_{mn}=\omega_{m}-\omega_{n}$. Meanwhile, other matrix elements of the type $\g{G}_{mmmm}(\tau_i-\tau_j)$ and $\g{G}_{mpmq}(\tau_i-\tau_j)$ with $m\not=p,q$ and $p\not=q$ need to be evaluated based on a specific model. This evaluation will be done in the next section based on a specific model of vibrational energy exchange taking place between two different molecules.

\section{ Molecular model undergoing intermolecular vibrational energy exchange}

\begin{figure}[h]
\begin{center}
\includegraphics[clip,scale=0.86,angle=0]{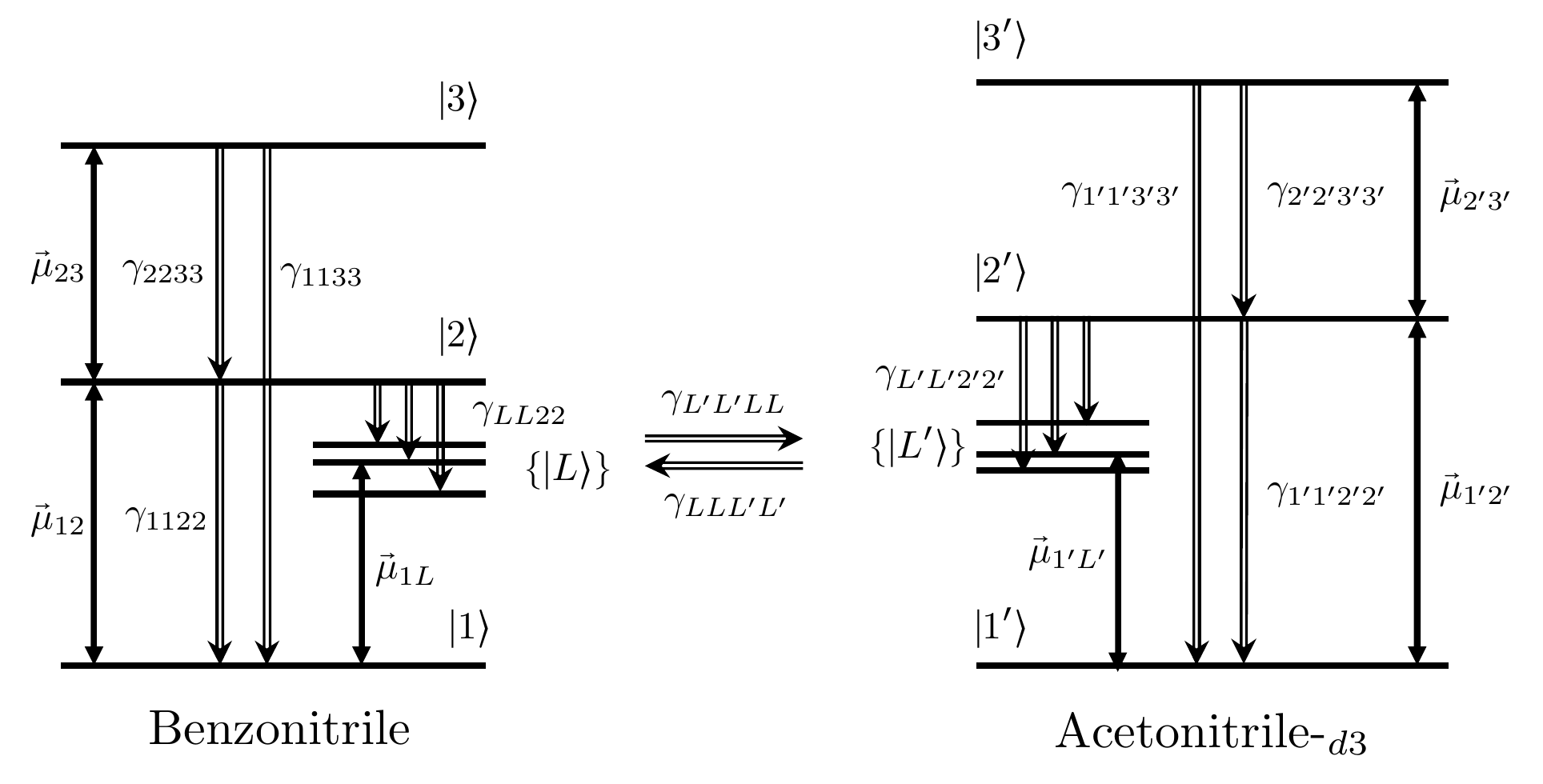} 
\end{center}
\caption{ Energy level scheme of the model system, the complex of acetonitrile-$_{\rm d3}$ (A) and benzonitrile (B). The various transition constants $\gamma_{nnpp}$ associated with the individual molecules are indicated by the single-headed arrows in the figure. The allowed dipole moments are indicated by double-headed arrows. The sets of states $\{\vert L\rangle\}$ and $\{\vert L'\rangle\}$ stand for the corresponding combination bands of their respective low frequency modes.}
\label{fig1}
\end{figure}

The molecular model undergoing intermolecular vibrational energy transfer between two molecules through a process of combination band absorption \cite{bb07,bb42}
and occurring among low frequency vibrational modes,  can be described by the Hamiltonian
\begin{eqnarray}
\g{H}=\sum_{i=a,b}\g{H}_{{\rm V}_i}+\g{H}_{{\rm L}_i}
\label{3.1}
\end{eqnarray}
where each individual molecular Hamiltonian $i=a,b$ is made of two parts, namely, an anharmonic vibrational mode $\g{H}_{{\rm V}_i}$ and a low frequency vibrational modes $\g{H}_{{\rm L}_i}$. The particular system studied in the following is the complex of acetonitrile-$_{\rm d3}$ (A) and benzonitrile (B), and the combination bands are most probably identified as the C-D bending mode in the deuterated methyl group of A and the benzene ring breathing mode in B \cite{bb21}. 
The relevant parameters of the individual molecules including energy levels and nonradiative transition rate constants, as well as the transition dipole moments $\vec{\mu}_{np}$ of the individual molecules with corresponding transition energies pertaining to the 2D-IR spectral range investigated, are shown in Fig.\ref{fig1}. Then, according to the sum rules
$\gamma_{pppp}=-\sum_{n\neq p}\gamma_{nnpp}$ and the definitions of the dephasing constants $\gamma_{npnp}=\gamma_{np}^{(d)}+(\gamma_{nnnn}+\gamma_{pppp})/2$ for the individual molecules, we can evaluate the various dynamical parameters of the total system. As usual, $\gamma_{pppp}$, $\gamma_{nnpp}$, $\gamma_{npnp}$ and $\gamma_{np}^{(d)}$ stand for the total decay rates, the transition constants, the dephasing and the pure dephasing constants of the individual molecules. For the total system, the corresponding quantities are denoted $\Gamma_{rrrr}$, $\Gamma_{rrss}$, $\Gamma_{rsrs}$ and $\Gamma_{rs}^{(d)}$, respectively. 
Notice that for small energy gaps, temperature effects can be efficient so that direct and reverse transition rate constants, related by detailed balance, are introduced. For the $\vert LL\rangle\rightleftarrows \vert L'L'\rangle$ transitions, we have the temperature dependence
\begin{eqnarray}
&&\gamma_{LLL'L'}=\gamma_{L'L'LL}\e^{\Delta E_{L'L}/kT}
\label{3.2}
\end{eqnarray}
and similar relation exists for the $\vert 22\rangle\rightleftarrows \vert LL\rangle$ and $\vert 2'2'\rangle\rightleftarrows \vert L'L'\rangle$ transitions.
\begin{figure}[h]
\begin{center}
\includegraphics[clip,scale=0.86,angle=0]{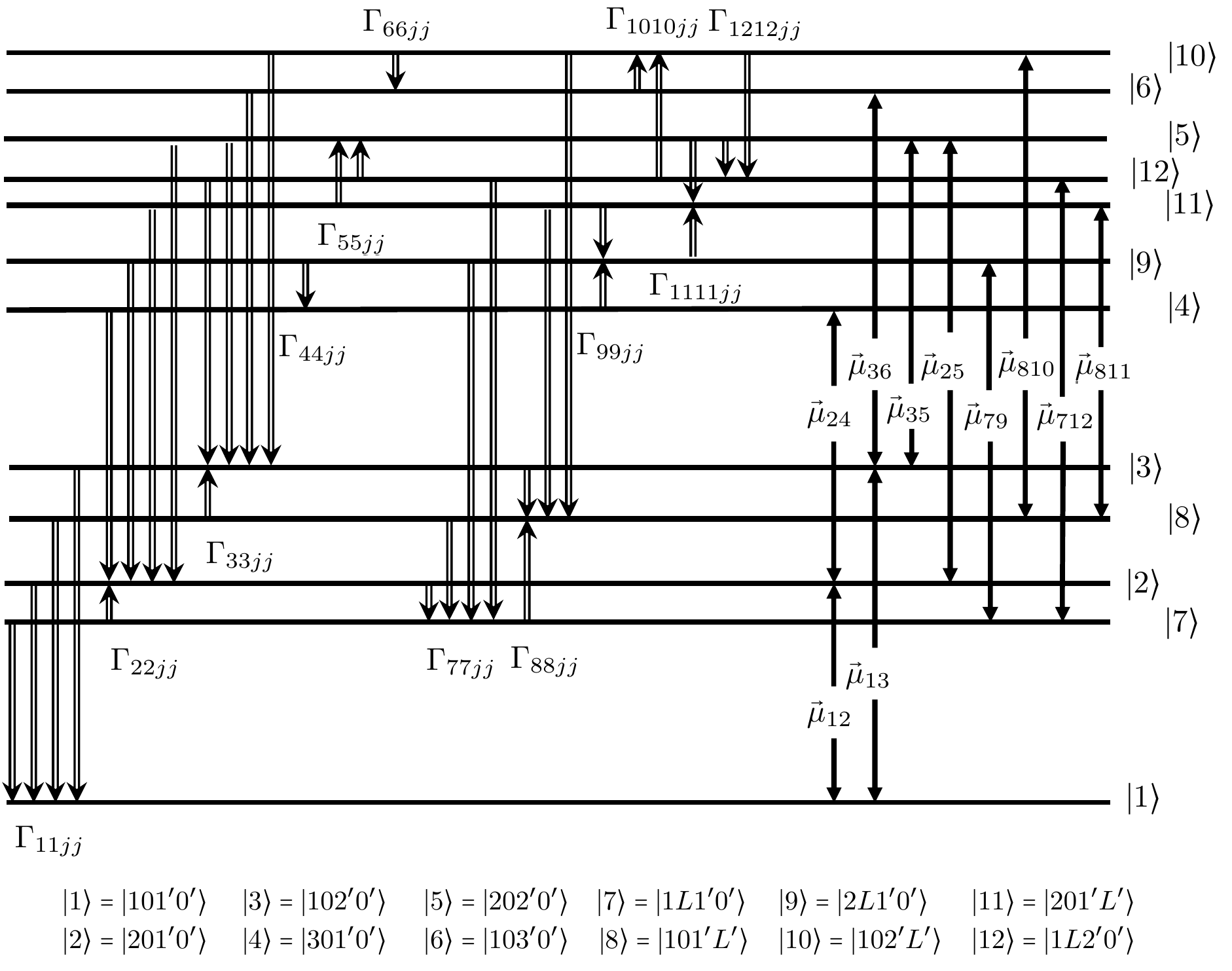} 
\end{center}
\caption{Energy level scheme of the total vibrational system previously discussed. The various transition constants $\Gamma_{iijj}$ and dipole moments associated to the total system are indicated using the same symbols as in Fig.\ref{fig1}. The set of states of the total vibrational system are defined at the bottom of the figure.}
\label{fig2}
\end{figure} 
We still have to introduce the coherence transfer constants $\Gamma_{mpmq}$. They introduce a coupling between the coherences associated to the transitions $\vert m\rangle \rightarrow \vert p\rangle$ and $\vert m\rangle \rightarrow \vert q\rangle$. The phenomenon of coherence transfer is now well established experimentally and has been observed in a wide range of systems including photodissociation of triatomic molecules \cite{bb43}, 
photobiological process in light harvesting compounds \cite{bb44}
or temperature-dependent spontaneous Raman spectra of high-frequency C-H vibrations in microwave rotational spectra \cite{bb45}. 
Moreover, it has also been considered theoretically \cite{bb18,bb46,bb47,bb48,bb49}.
In the present system, the more important consequence of coherence transfer, especially between closely-lying vibrational energy levels, is the ability of observing dark transitions appearing as extra-peaks in the 2D-IR spectra.

All of the vibrational states of the total molecular system are described at the bottom of Fig.\ref{fig2} where their corresponding dipole moments and transition constants are presented. From the laser-molecule interactions among the set of states ranging from $\vert 1\rangle$ to $\vert 12 \rangle$ and their corresponding relaxation constants, the various pathways in the total Liouvillian space participating in the dynamical evolution can be determined. As usual, only pathways satisfying the rotating wave approximation (RWA) will be retained. The individual pathways are given in Appendix A and their corresponding contributions to the various density matrix elements can be obtained from  Eq.(\ref{2.2}), from the knowledge of the coherence evolution previously established, and also from the population and coherence transfer evolutions that need to be evaluated. This has to be done according to the specific vibrational model of energy transfer.

The evolution of the populations is straightforwardly deduced from the zeroth-order Liouville equation 
\begin{eqnarray}
\frac{\partial \g{\rho}_{mm}(t)}{\partial t}&=& -\g{\Gamma}_{mmnn}\g{\rho}_{nn}(t).
\label{3.3}
\end{eqnarray}
where, in the present model, the relaxation Liouvillian takes the form
\begin{eqnarray}
\setcounter{MaxMatrixCols}{12}
&&\g{\Gamma}=
\begin{pmatrix}
\ssc{0}\;&\;\ssc{{\gamma}_{1122}}\;&\;\ssc{{\gamma}_{1133}}\;&\;\ssc{0}\;&\;\ssc{0}\;&\;\ssc{0}\;&\;\ssc{{\gamma}_{1177}}\;&\;\ssc{{\gamma}_{1188}}\;&\;\ssc{0}\;&\;\ssc{0}\;&\;\ssc{0}\;&\;\ssc{0} \\
\ssc{0}\;&\;\ssc{\Gamma[2,2,2,2]} \;&\;\ssc{0}\;&\;\ssc{{\gamma}_{2233}}\;&\;\ssc{{\gamma}_{1'1'2'2'}}\;&\;\ssc{0}\;&\;\ssc{{\gamma}_{22LL}}\;&\;\ssc{0}\;&\;\ssc{{\gamma}_{11LL}}\;&\;\ssc{0}\;&\;\ssc{{\gamma}_{1'1'L'L'}}\;\ssc{0}\\
\ssc{0}\;&\;\ssc{0}\;&\;\ssc{\Gamma[3,3,3,3]}\;&\;\ssc{0}\;&\;\ssc{{\gamma}_{1122}}\;&\;\ssc{{\gamma}_{2'2'3'3'}}\;&\;\ssc{0}\;&\;\ssc{{\gamma}_{2'2'L'L'}}\;&\;\ssc{0}\;&\;\ssc{{\gamma}_{1'1'L'L'}}\;&\;\ssc{0}\;&\;\ssc{{\gamma}_{11LL}}\\
\ssc{0}\;&\;\ssc{0}\;&\;\ssc{0}\;&\;\ssc{\Gamma[4,4,4,4]}\;&\;\ssc{0}\;&\;\ssc{0}\;&\;\ssc{0}\;&\;\ssc{0}\;&\;\ssc{{\gamma}_{22LL}}\;&\;\ssc{0}\;&\;\ssc{0}\;&\;\ssc{0}\\ 
\ssc{0}\;&\;\ssc{0}\;&\;\ssc{0}\;&\;\ssc{0}\;&\;\ssc{\Gamma[5,5,5,5]}\;&\;\ssc{0}\;&\;\ssc{0}\;&\;\ssc{0}\;&\;\ssc{{\gamma}_{22LL}}\;&\;\ssc{0}\;&\;\ssc{{\gamma}_{2'2'L'L'}}\;&\;\ssc{{\gamma}_{22LL}}\\
\ssc{0}\;&\;\ssc{0}\;&\;\ssc{0}\;&\;\ssc{0}\;&\;\ssc{0}\;&\;\ssc{\Gamma[6,6,6,6]}\;&\;\ssc{0}\;&\;\ssc{0}\;&\;\ssc{{\gamma}_{22LL}}\;&\;\ssc{{\gamma}_{2'2'L'L'}}\;&\;\ssc{0}\;&\;\ssc{0}\\
\ssc{0}\;&\;\ssc{{\gamma}_{LL22}}\;&\;\ssc{0}\;&\;\ssc{0}\;&\;\ssc{0}\;&\;\ssc{0}\;&\;\ssc{\Gamma[7,7,7,7]}\;&\;\ssc{{\gamma}_{LLL'L'}}\;&\;\ssc{{\gamma}_{1122}}\;&\;\ssc{0}\;&\;\ssc{0}\;&\;\ssc{{\gamma}_{1'1'2'2'}}\\ 
\ssc{0}\;&\;\ssc{0}\;&\;\ssc{{\gamma}_{L'L'2'2'}}\;&\;\ssc{0}\;&\;\ssc{0}\;&\;\ssc{0}\;&\;\ssc{{\gamma}_{L'L'LL}}\;&\;\ssc{\Gamma[8,8,8,8]}\;&\;\ssc{0}\;&\;\ssc{{\gamma}_{1'1'2'2'}}\;&\;\ssc{{\gamma}_{1122}}\;&\;\ssc{0}\\ 
\ssc{0}\;&\;\ssc{0}\;&\;\ssc{0}\;&\;\ssc{{\gamma}_{LL22}}\;&\;\ssc{0}\;&\;\ssc{0}\;&\;\ssc{0}\;&\;\ssc{0}\;&\;\ssc{\Gamma[9,9,9,9]}\;&\;\ssc{0}\;&\;\ssc{{\gamma}_{LLL'L'}}\;&\;\ssc{0}\\ 
\ssc{0}\;&\;\ssc{0}\;&\;\ssc{0}\;&\;\ssc{0}\;&\;\ssc{0}\;&\;\ssc{{\gamma}_{L'L'2'2'}}\;&\;\ssc{0}\;&\;\ssc{0}\;&\;\ssc{0}\;&\;\ssc{\Gamma[10,10,10,10]}\;&\;\ssc{0}\;&\;\ssc{{\gamma}_{L'L'LL}}\\ 
\ssc{0}\;&\;\ssc{0}\;&\;\ssc{0}\;&\;\ssc{0}\;&\;\ssc{{\gamma}_{L'L'2'2'}}\;&\;\ssc{0}\;&\;\ssc{0}\;&\;\ssc{0}\;&\;\ssc{{\gamma}_{L'L'LL}}\;&\;\ssc{0}\;&\;\ssc{\Gamma[11,11,11,11]}\;&\;\ssc{0}\\
\ssc{0}\;&\;\ssc{0}\;&\;\ssc{0}\;&\;\ssc{0}\;&\;\ssc{{\gamma}_{LL22}}\;&\;\ssc{0}\;&\;\ssc{0}\;&\;\ssc{0}\;&\;\ssc{0}\;&\;\ssc{{\gamma}_{LLL'L'}}\;&\;\ssc{0}\;&\;\ssc{\Gamma[12,12,12,12]}
\end{pmatrix} \nonumber\\
&&
\label{3.4}
\end{eqnarray}\setcounter{MaxMatrixCols}{10}
Besides, from our model, the total decay rates are related to the transition rates constants by the expressions given in Table \ref{1} 

{
\begin{table}[h]
\begin{center}
\begin{tabularx}{\textwidth}{XXX}
\hline \hline
$\ssc{\Gamma[2,2,2,2]=0}$&
$\ssc{\Gamma[2,2,2,2]=-\sum_{u=1,7}\Gamma[u,u,2,2]}$&
$\ssc{\Gamma[3,3,3,3]=-\sum_{u=1,8}\Gamma[u,u,3,3]}$  \\
\hline
$\ssc{\Gamma[4,4,4,4]=-\sum_{u=2,9}\Gamma[u,u,4,4]}$&
$\ssc{\Gamma[5,5,5,5]=-\sum_{u=2,3,11,12}\Gamma[u,u,5,5]}$&
$\ssc{\Gamma[6,6,6,6]=-\sum_{u=3,10}\Gamma[u,u,6,6]}$\\
\hline
$\ssc{\Gamma[7,7,7,7]=-\sum_{u=1,2,8}\Gamma[u,u,7,7]}$&
$\ssc{\Gamma[8,8,8,8]=-\sum_{u=1,3,7}\Gamma[u,u,8,8]}$&
$\ssc{\Gamma[9,9,9,9]=-\sum_{u=2,4,7,11}\Gamma[u,u,9,9]}$\\
\hline
$\ssc{\Gamma[10,10,10,10]=-\sum_{u=6,8,3,12}\Gamma[u,u,10,10]}$&
$\ssc{\Gamma[11,11,11,11]=-\sum_{u=2,5,8,9}\Gamma[u,u,11,11]}$&
$\ssc{\Gamma[12,12,12,12]=-\sum_{u=3,5,7,10}\Gamma[u,u,12,12]}$\\
\hline \hline
\end{tabularx}
\caption{Relations between total decay rates and transition rate constants.}
\label{1}
\end{center}
\end{table}}
Then, by identifying the integral representation of the density matrix with the definition of the population evolution Liouvillian, it is found that 
\begin{eqnarray}
\g{\rho}_{mm}(t)=\frac{1}{2\pi i}\int_{-\infty+i\epsilon}^{\infty+i\epsilon}ds\, \e^{st}\bigl[s\g{I}+\g{\Gamma}\bigr]^{-1}_{mmnn}\g{\rho}_{nn}(t_0)
=\sum_{m}\g{G}_{mmnn}(t-t_0)\g{\rho}_{nn}(t_0)
\label{3.5}
\end{eqnarray}
and the evolution Liouvillians in the population subspace are deduced from the diagonalization of $\bigl[s\g{I}+\g{\Gamma}\bigr]^{-1}$. They correspond to $\g{G}_{1111}(t)=1$ for the ground state and all the other matrix elements can be expressed in terms of the eigenvalues $\lambda_{\alpha}$ and the components of their corresponding eigenvectors $W(\lambda_{\alpha})$ of $\bigl[s\g{I}+\g{\Gamma}\bigr]^{-1}$, so that
\begin{eqnarray}
\g{G}_{mmnn}(t)=\sum_{\alpha}W_{mmnn}(\lambda_{\alpha})e^{\lambda_{\alpha}t}.
\label{3.6}
\end{eqnarray}
%
\begin{figure}[h]
\begin{center}
\includegraphics[clip,scale=0.86,angle=0]{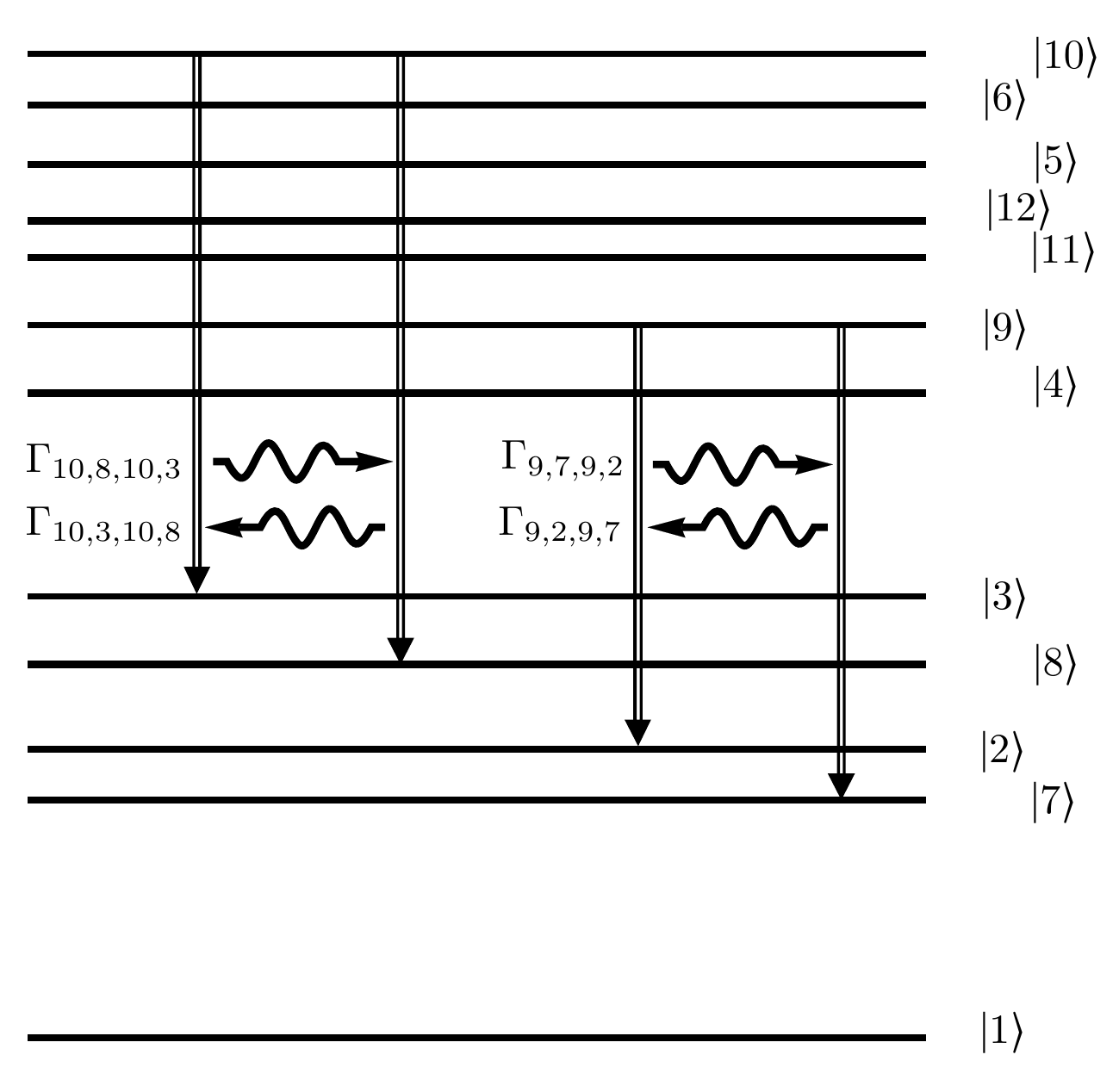} 
\end{center}
\caption{The direct and reverse processes of coherence transfer between transitions $\vert 9\rangle \rightarrow \vert 7\rangle$ and $\vert 9\rangle \rightarrow \vert 2\rangle$. The same processes occurs between transitions $\vert 10\rangle \rightarrow \vert 8\rangle$ and $\vert 10\rangle \rightarrow \vert 3\rangle$ as illustrated in this figure.} 
\label{fig3}
\end{figure} 
%
It remains to calculate the evolution Liouvillian associated with the coherence transfer processes depicted in Fig.\ref{fig3}. To this end, we have to diagonalize the evolution Liouvillian in the coherence subspace. Only, the coherence subspace  $\vert i,j\not=i\rangle\!\!\!\rangle$ coupled by the coherence transfer constants need to be considered. As shown in Fig.\ref{fig3}, they are the $\vert 9\rangle \rightarrow \vert 7 \rangle$ and $\vert 9 \rangle \rightarrow \vert 2 \rangle$ transitions in the first case, as well as the $\vert 10\rangle \rightarrow \vert 8 \rangle$ and $\vert 10 \rangle \rightarrow \vert 3 \rangle$ transitions in the second case. Therefore, in the Liouvillian subspace where coherence transfer takes place, we have for $m\not=p,q$ and $p\not=q$
\begin{multline}
\g{G}_{mpmq}(t)=\frac{1}{2\pi i}\int_{-\infty+i\epsilon}^{\infty+i\epsilon}ds\, \e^{st} \\
\times\begin{pmatrix}
s+i\omega_{92}+\Gamma_{9292}&\Gamma_{9792}&0&0 \\
\Gamma_{9297}&s+i\omega_{97}+\Gamma_{9797}&0&0 \\  
0&0&s+i\omega_{103}+\Gamma_{103103}&\Gamma_{108103} \\  
0&0&\Gamma_{103108}&s+i\omega_{108}+\Gamma_{108108}
\end{pmatrix}_{mpmq}
\label{3.7}
\end{multline}
Now, we have all the required quantities to perform the analytical evaluation and subsequent numerical simulations of the 2D-IR spectra for a system undergoing energy transfer. Eq.(3.7) indicates that the coherence transfer processes are described by the coherence transfer constants $\Gamma_{9792}$ and $\Gamma_{108103}$ involving the dark states $\vert L\rangle$ and $\vert L'\rangle$ of the benzonitrile and acetonitrile-$_{d3}$ molecules, respectively. Therefore, by coupling a dark state to an optically active state, it is possible to visualize on the two-dimensional spectrum the resonance associated with the energy transfer.

\section{ Simulation of 2D-IR spectra in presence of energy transfer}

From the previous sections, the expressions of the Liouville evolution operators in the population, coherence and coherence transfer subspaces can be deduced straightforwardly and evaluated for all the matrix elements appearing in the various pathways given in Appendix A. Once their expressions, taken for a given path, are introduced in Eq.(\ref{2.2}) and identified with the density matrix element (\ref{2.9}) using the formal expression (\ref{2.10}) for $\g{R}_{n_p,ij}(\tau_1,\tau_2,\tau_3,t)$, the various quantities $A_{n_p,r,q,p}$, $B_{n_p,r,q}$, $C_{n_p,r}$, $K_{n_p,r,q,p}$ and $Q_{n_p,r,q,p}$ can be obtained for each individual pathway. The final step to obtain the analytical expression of the 2 D-IR spectrum is the triple time integration plus the double Fourier transform over $\tau$  and $t$. They need to be performed for each field ordering, as well as for the two definitions of $\tau$ according to the sign of $T_b-T_a$. Their expressions are a little bit cumbersome. For this reason, they will not be presented here. As an example, we give the expression corresponding to the field ordering $c\leftarrow b\leftarrow a$. Other ones, can be evaluated as well.         


At this stage, it is necessary to introduce all of the numerical values required to perform the numerical simulations from the analytical results describing the vibrational energy exchange between the acetonitrile-$_{\rm d3}$ and benzonitrile molecules, chosen as the model system. 
{\scriptsize
\begin{table}[h]
\begin{center}
\begin{tabular}{|r|r|r|r|} 
\hline \hline
\multicolumn{4}{|l|}{Benzonitrile molecule}\\
\hline
$\ssc{\omega_{1}=0}$&$\ssc{\omega_{2}=2230}$&$\ssc{\omega_{3}=4439}$&$\ssc{\omega_L=2224}$  \\
\hline
\hline
\multicolumn{4}{|l|}{Acetonitrile-$_{d3}$ molecule}\\
\hline
$\ssc{\omega_{1'}=0}$&$\ssc{\omega_{2'}=2263}$&$\ssc{\omega_{3'}=4508}$&$\ssc{\omega_{L'}=2254}$  \\
\hline \hline
\end{tabular} 
\caption{All of the vibrational energies are expressed in units of ${\rm cm}^{-1}$.} 
\label{2}
\end{center}
\end{table}}
For simplicity, only one combination band state per molecule will be retained in the following simulations. The energies of the acetonitrile-$_{d3}$ and benzonitrile molecules are given in Table \ref{2}.
{\scriptsize
\begin{table}[h]
\begin{center}
\begin{tabular}{|l|l|l|l|} 
\hline \hline 
Total decay rates&Transition constants&Pure dephasings&Coherence transfer constants\\
\hline\hline
$\ssc{\gamma_{1111}^{-1}=\infty}$&
$\ssc{\gamma_{1133}^{-1}=\infty}$&
$\ssc{\gamma_{12}^{(d)-1}=2}$&
$\ssc{\Gamma_{9297}^{-1}=3}$  \\
\hline
$\ssc{\gamma_{1'1'1'1'}^{-1}=\infty}$&
$\ssc{\gamma_{1'1'3'3'}^{-1}=\infty}$&
$\ssc{\gamma_{1'2'}^{(d)-1}=2}$&
$\ssc{\Gamma_{103108}^{-1}=2}$\\
\hline
$\ssc{\gamma_{2222}^{-1}=3.1}$&
$\ssc{\gamma_{LL22}^{-1}=3.9}$&
$\ssc{\gamma_{23}^{(d-1)}=3}$&
\\
\hline
$\ssc{\gamma_{2'2'2'2'}^{-1}=6.1}$&
$\ssc{\gamma_{L'L'LL}^{-1}=40}$&
$\ssc{\gamma_{2'3'}^{(d)-1}=3}$&
\\
\hline
$\ssc{\gamma_{3333}^{-1}=5}$&
$\ssc{\gamma_{11LL}^{-1}=40}$&
$\ssc{\gamma_{13}^{(d)-1}=4}$&
\\
\hline
$\ssc{\gamma_{3'3'3'3'}^{-1}=5}$&
$\ssc{\gamma_{1'1'L'L'}^{-1}=31}$&
$\ssc{\gamma_{1'3'}^{(d)-1}=4}$&
\\
\hline
&
$\ssc{\gamma_{L'L'2'2'}^{-1}=7.6}$&
$\ssc{\gamma_{1L}^{(d)-1}=2}$&
\\
\hline 
&
$\ssc{}$&
$\ssc{\gamma_{1'L'}^{(d)-1}=2}$&
\\
\hline
&
&
$\ssc{\gamma_{AB}^{(d)-1}=5}$&
\\ 
\hline \hline
\end{tabular} 
\caption{Numerical values of the total decay rates, transition rate constants and pure dephasing constants of the individual molecules. All of the numerical values are given in units of ps.} 
\label{3}
\end{center}
\end{table}}
\begin{figure}[h]
\centering
{\subfigcapskip=-20pt\subfigure{\includegraphics[width=0.38\textwidth]{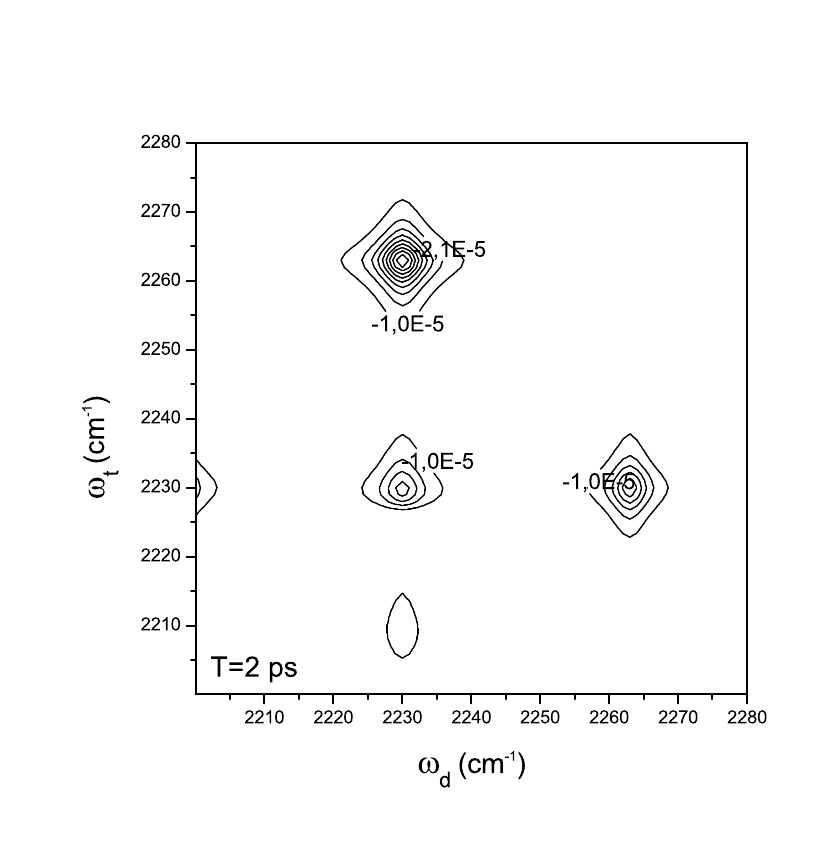}}\hskip -1.8truecm                
\subfigure{\includegraphics[width=0.38\textwidth]{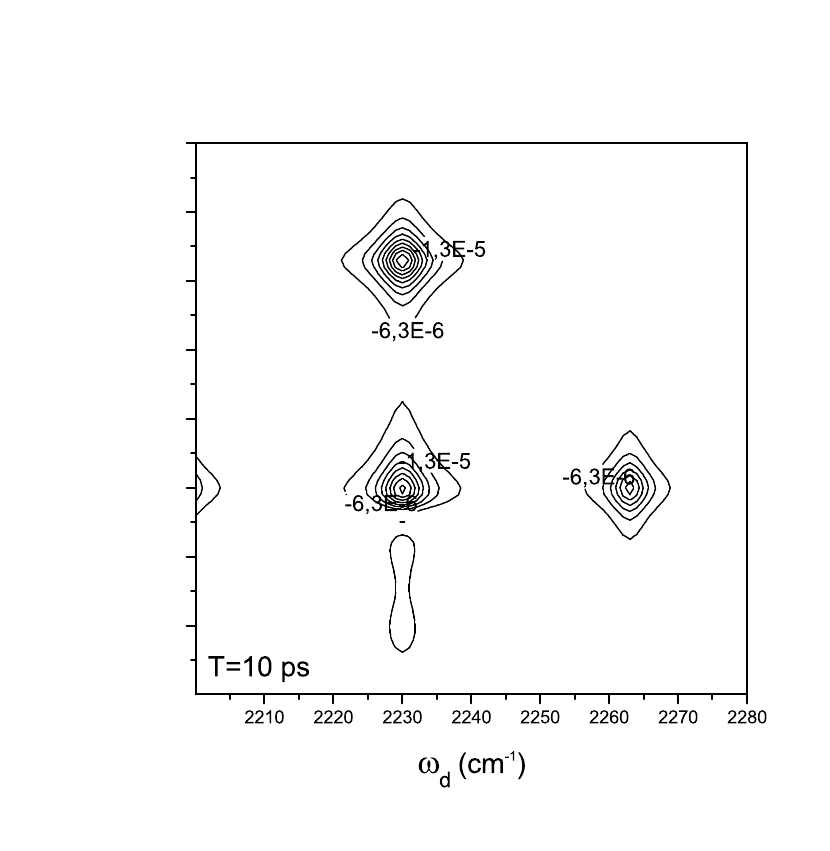}}\hskip -1.8truecm
\subfigure{\includegraphics[width=0.38\textwidth]{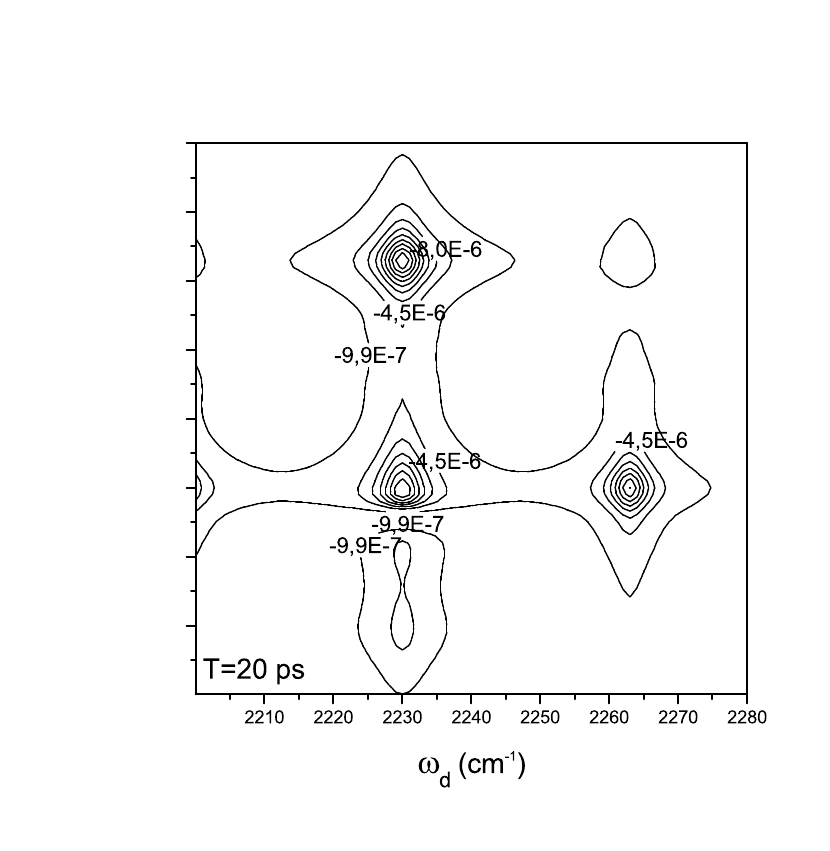}}\hskip -1.8truecm \vskip -1truecm
\subfigure{\includegraphics[width=0.38\textwidth]{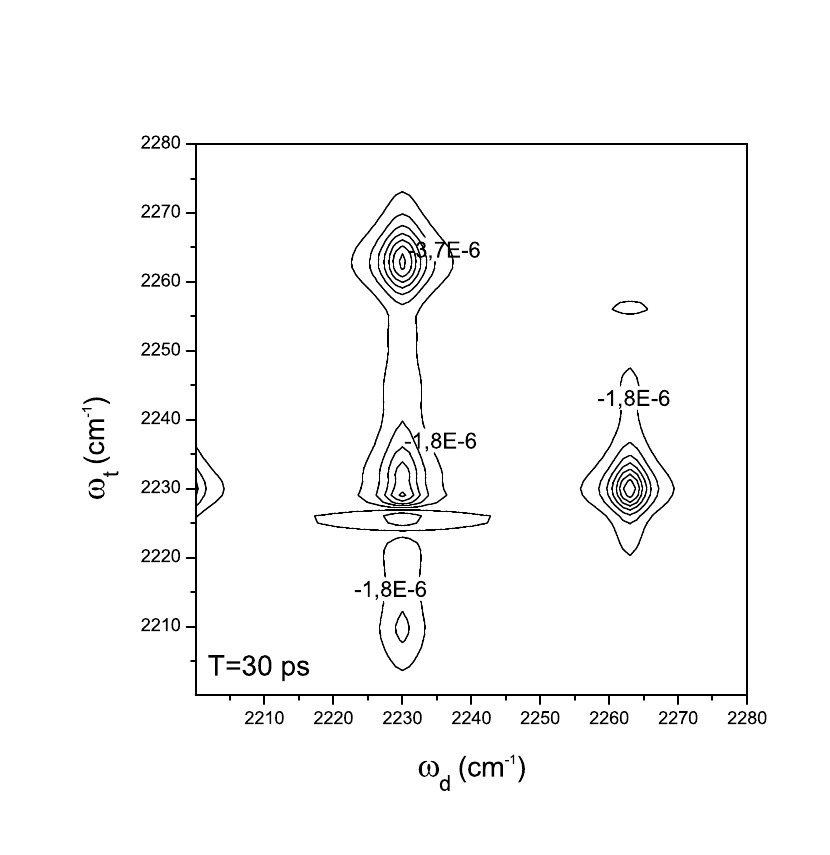}}\hskip -1.8truecm 
\subfigure{\includegraphics[width=0.38\textwidth]{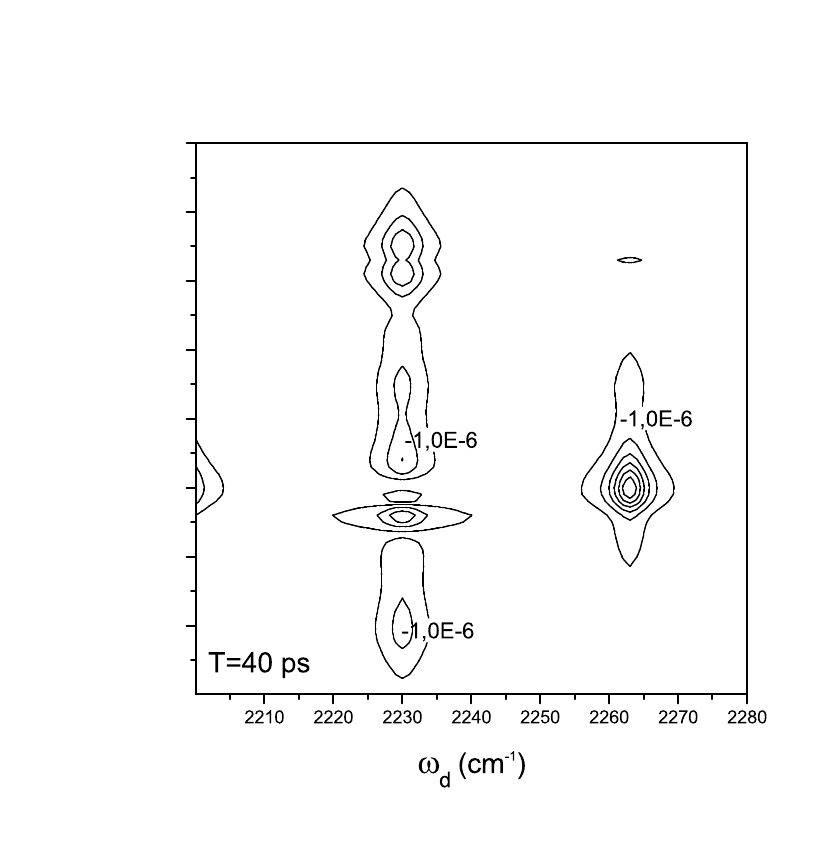}}}
\caption{Two-dimensional spectra obtained for different waiting times T between the two last interacting laser pulses, say $T=min(|T_a|,|T_b|)$ as defined in the text. The phase of the local field oscillator used for heterodyne detection is chosen, here, to be $\Psi=0$.} 
\label{fig4}
\end{figure}
\begin{figure}[ht!]
\centering
{\subfigcapskip=-20pt\subfigure{\includegraphics[width=0.38\textwidth]{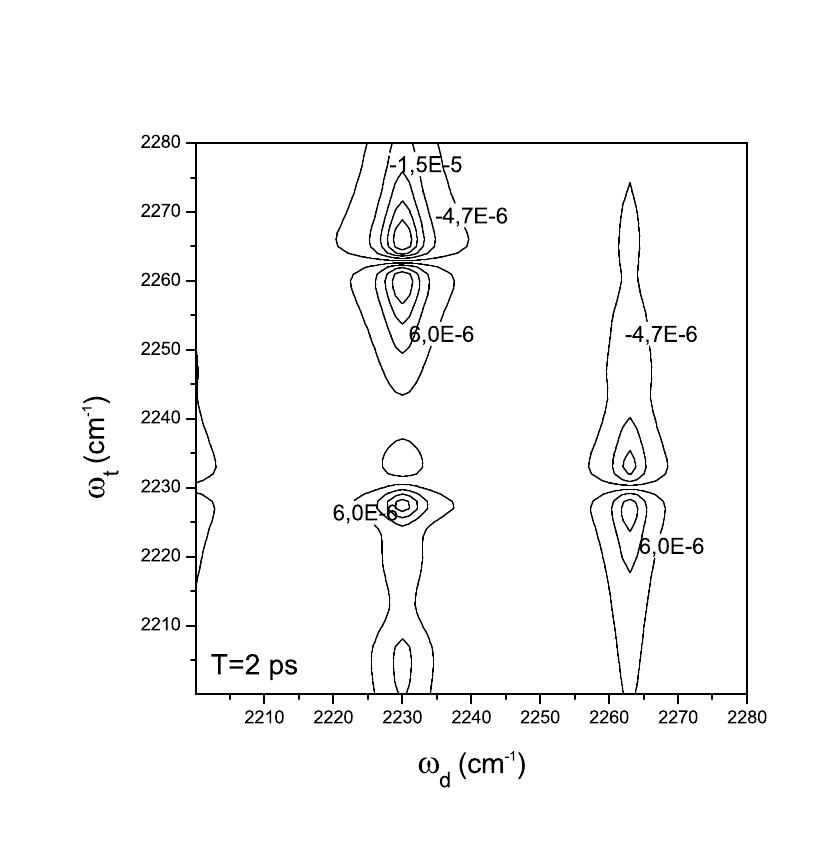}}\hskip -1.8truecm                
\subfigure{\includegraphics[width=0.38\textwidth]{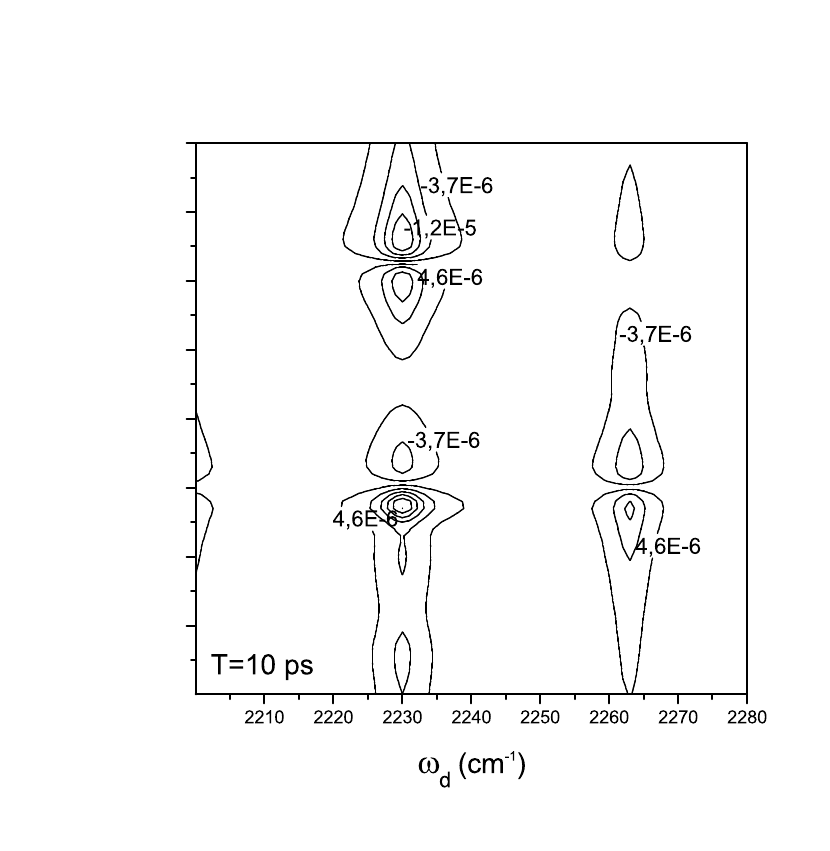}}\hskip -1.8truecm
\subfigure{\includegraphics[width=0.38\textwidth]{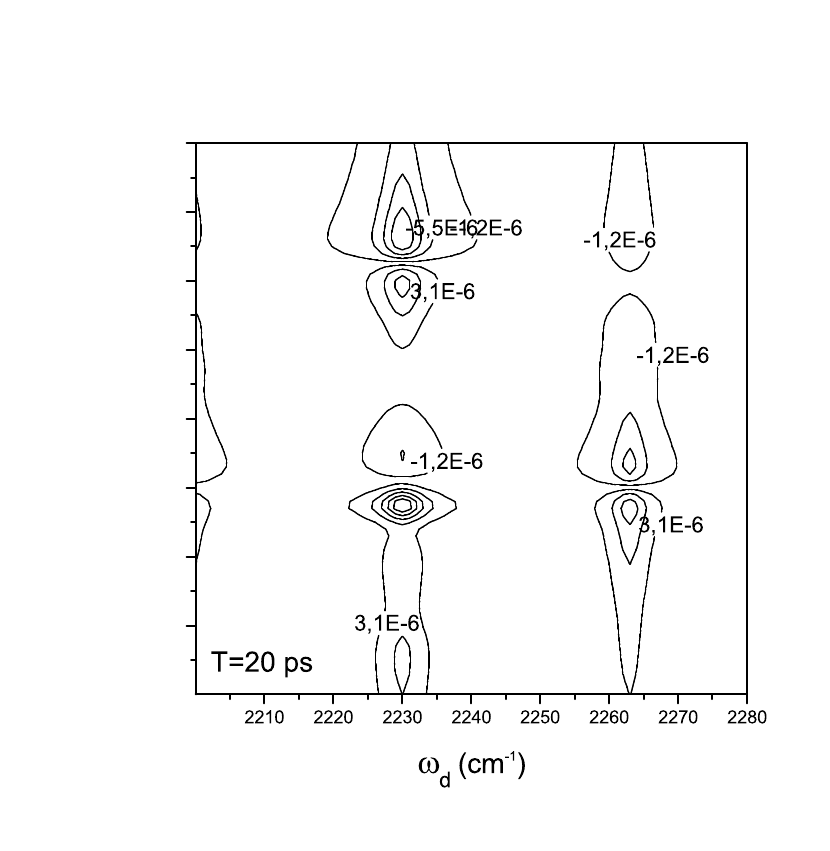}}\hskip -1.8truecm \vskip -1truecm
\subfigure{\includegraphics[width=0.38\textwidth]{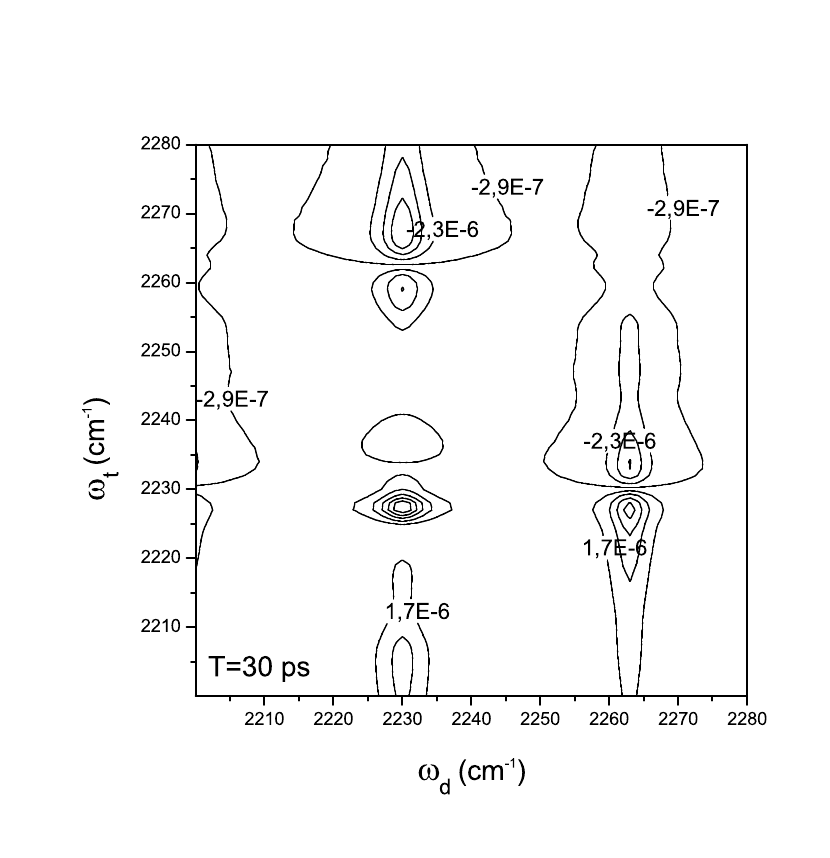}}\hskip -1.8truecm 
\subfigure{\includegraphics[width=0.38\textwidth]{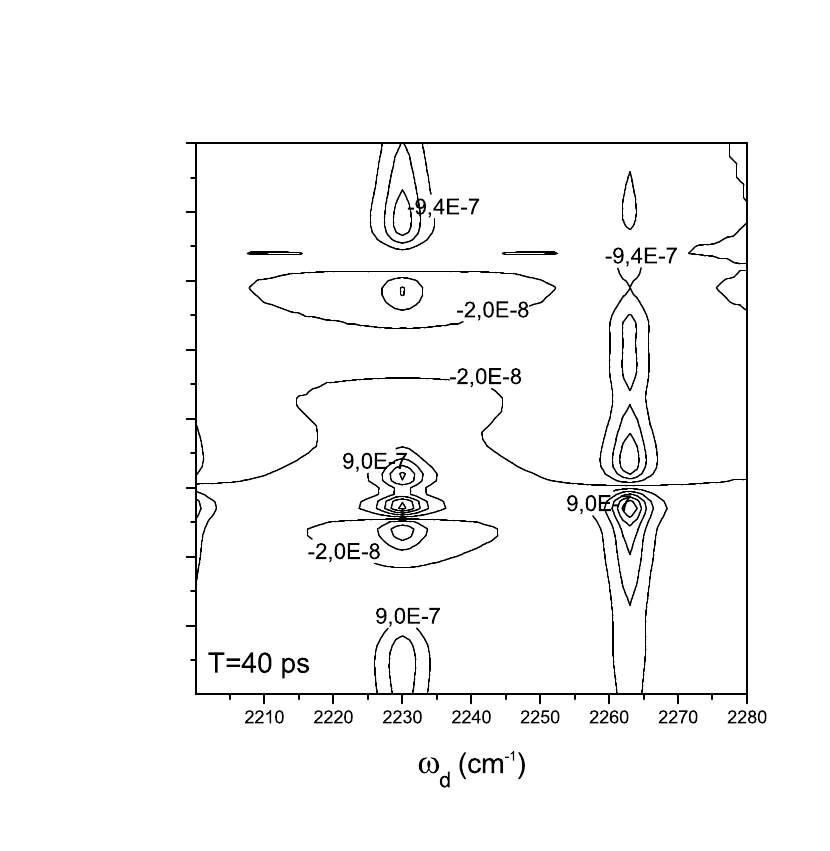}}}
\caption{The same variations as those shown in Fig.\ref{fig4} are presented. The difference is in that here the phase of the local field oscillator is fixed at $\Psi=\pi/2$. } 
\label{fig5}
\end{figure}
\pagebreak[5]
Next, the relaxation constants of the individual molecules, as well as the coherence transfer constants of the total system, are shown in Table \ref{3}.
Besides, the dipole moments are $\mu_{12}=0.9$ and $\mu_{23}=0.7$ for the acetonitrile-$_{\rm d3}$ and $\mu_{12}=1$ and $\mu_{23}=0.7$ for the benzonitrile molecules. 


In Fig.\ref{fig4}, we present the 2D-IR spectra of the acetonitrile-$_{d3}$ and benzonitrile molecules when the phase of the local oscillator is fixed at $\Psi=0$. The various panels correspond to different waiting times $T$, as defined in Section 2, and corresponding to the time of delay between the two last interacting laser pulses. Their values range from $T=2$ps to $T=40$ps, and are indicated at the bottom of each panels. In the first panel, the molecular system has been excited and then tested after a period of time short enough so that energy transfer cannot take place because it occurs on a much longer period. In that case, only one diagonal peak corresponding to the excitation of the benzonitrile molecule alone located at ($2230\,{\rm cm}^{-1}$, $2230\,{\rm cm}^{-1}$) can be observed. The corresponding diagonal peak of the acetonitrile-$_{d3}$ molecule is too small to appear for the particular physical parameters chosen here.
However, the cross peaks involving simultaneous excitations of both molecules appear at ($2230\,{\rm cm}^{-1}$, $2263\,{\rm cm}^{-1}$) and vice versa. For increasing values of the waiting time, say $T=10\,ps$, we observe some changes in the spectrum related to the beginning of the energy transfer process can be observed. However, at this stage, since the peaks are quite close to each other, they overlap significantly and their changes depend on many physical parameters so that no conclusive observation can be made about the energy transfer process. With longer waiting times, the peak at ($2230\,{\rm cm}^{-1}$, $2224\,{\rm cm}^{-1}$), which is the signature of the energy transfer process, becomes important enough so that the minor changes at early time are dominated by the rise of this peak. Finally, for the waiting times $T=30$ps and $T=40$ps, the peak resulting from the coherence transfer involving the dark combination state $\vert L\rangle$ and the optically active state $\vert 2\rangle$ of the benzonitrile molecule is clearly identified on the 2D-IR spectrum. 


The last numerical simulations are shown in Fig.\ref{fig5}. There, the 2D-IR spectra obtained using the same physical parameters as in Fig.\ref{fig4} are shown, except that the phase of the local field oscillator is chosen to be $\Psi=\pi/2$. It is important to mention that the observation of the peak at ($2230\,{\rm cm}^{-1}$, $2224\,{\rm cm}^{-1}$) is now much more difficult. As previously discussed for the acetonitrile-$_{d3}$ and benzonitrile molecules, the peaks are quite close to each other and their energy gap is of the same order of magnitude as their resonance linewidths. The $\Psi=\pi/2$ heterodyne detection effectively picks up the imaginary part of the polarization, the frequency dependence of which is dispersive.  
Due to the general shape of the dispersion curves, the peak associated with the energy transfer overlaps much more with the lower part of the diagonal at ($2230\,{\rm cm}^{-1}$, $2230\,{\rm cm}^{-1}$). Then, the resulting distorsion is still more enhanced because of the proximity of both the diagonal and the energy transfer peaks. This is why the observation of the energy transfer peak is more difficult for the local oscillator phase $\Psi=\pi/2$ than for $\Psi=0$.


\section{ Conclusion}

In this work, an energy exchange process observed by two-dimensional spectroscopy has been described analytically . The observation of this process is possible because of the coherence transfer processes taking place among combination bands of the low frequency modes associated with the different molecules participating in the energy exchange process. The influence of this coherence transfer on the dynamics of the total molecular system is of particular interest because it reveals resonances associated with these dark modes that could not be observed otherwise. However, even for small coherence transfer rate constants, the resonance are slightly shifted. For the specific vibrational model considered here, the various transition frequencies participating in the two-dimensional spectra are close to each other, therefore the structure of the peak, which is a signature of the energy exchange, is very sensitive to the magnitude of coherence transfer constants. It will be interesting to get a more detailed description of these coherence transfer constants to get better insights of their influence on the energy transfer process.

\begin{acknowledgments}
This work was supported by the Academia Sinica and by the National Science Council of Taiwan (grant NSC 100-2113-M-001-006-MY2). One of us, (A.A. Villaeys), is indebted to Dr. K.K. Liang for his kind hospitality during his stay in Nankang. 
\end{acknowledgments}


\appendix
\section{Pathways involved}
All of the pathways can be classified according to their particular molecular-field interaction orderings. Each table listed below corresponds to one particular ordering and only pathways satisfying the RWA approximation are retained. The first table corresponds to
\begin{center}
{\scriptsize
\begin{longtable}{c@{\phantom{---}}cccccccc}
\hline \hline
Path&$\!\scr{\g\rho(t)}\!$&$\!\scr{\g{G}(t-\tau_3)}\!$&$\!\scr{\g{L}_{v[p]}(\tau_3)}\!$&$\!\scr{\g{G}(\tau_3-\tau_2)}\!$&$\!\scr{\g{L}_{v[q]}(\tau_2)}\!$&$\!\scr{\g{G}((\tau_2-\tau_1))}\!$ &$\!\scr{\g{L}_{v[a(b)]}(\tau_1)}\!$&$\!\scr{\g{\rho}(t_0)}\!$\\
\hline
\hline
$\scr{1}$&$\scr{42}$&$\scr{4242}$&$\scr{4222}^{(+)}$&$\scr{2222}$&$\scr{2212}^{(+)}$&$\scr{1212}$&$\scr{1211}^{(-)}$&$\scr{11}$\\
\hline
$\scr{2}$&$\scr{52}$&$\scr{5252}$&$\scr{5222}^{(+)}$&&&&&\\
\hline
$\scr{3}$&$\scr{21}$&$\scr{2121}$&$\scr{2122}^{(+)}$&&&&&\\
\hline
$\scr{4}$&$\scr{21}$&$\scr{2121}$&$\scr{2111}^{(+)}$&$\scr{1122}$&&&&\\
\hline
$\scr{5}$&$\scr{31}$&$\scr{3131}$&$\scr{3111}^{(+)}$&&&&&\\
\hline
$\scr{6}$&$\scr{53}$&$\scr{5353}$&$\scr{5333}^{(+)}$&$\scr{3322}$&&&&\\
\hline
$\scr{7}$&$\scr{63}$&$\scr{6363}$&$\scr{6333}^{(+)}$&&&&&\\
\hline
$\scr{8}$&$\scr{31}$&$\scr{3131}$&$\scr{3133}^{(+)}$&&&&&\\
\hline
$\scr{9}$&$\scr{97}$&$\scr{9797}$&$\scr{9777}^{(+)}$&$\scr{7722}$&&&&\\
\hline
$\scr{10}$&$\scr{92}$&$\scr{9297}$&$\scr{9777}^{(+)}$&$\scr{7722}$&&&&\\
\hline
$\scr{11}$&$\scr{127}$&$\scr{127127}$&$\scr{12777}^{(+)}$&&&&&\\
\hline
$\scr{12}$&$\scr{118}$&$\scr{118118}$&$\scr{11888}^{(+)}$&$\scr{8822}$&&&&\\
\hline
$\scr{13}$&$\scr{108}$&$\scr{108108}$&$\scr{10888}^{(+)}$&&&&&\\
\hline
$\scr{14}$&$\scr{103}$&$\scr{103108}$&$\scr{10888}^{(+)}$&&&&&\\
\hline
$\scr{15}$&$\scr{52}$&$\scr{5252}$&$\scr{5232}^{(+)}$&$\scr{3232}$&$\scr{3212}^{(+)}$&&&\\
\hline
$\scr{16}$&$\scr{31}$&$\scr{3131}$&$\scr{3132}^{(+)}$&&&&&\\
\hline
$\scr{17}$&$\scr{21}$&$\scr{2121}$&$\scr{2111}^{(+)}$&$\scr{1111}$&$\scr{1112}^{(+)}$&&&\\
\hline
$\scr{18}$&$\scr{31}$&$\scr{3131}$&$\scr{3111}^{(+)}$&&&&&\\
\hline
$\scr{19}$&$\scr{53}$&$\scr{5353}$&$\scr{5333}^{(+)}$&$\scr{3333}$&$\scr{3313}^{(+)}$&$\scr{1313}$&$\scr{1311}^{(-)}$&\\
\hline
$\scr{20}$&$\scr{63}$&$\scr{6363}$&$\scr{6333}^{(+)}$&&&&&\\
\hline
$\scr{21}$&$\scr{31}$&$\scr{3131}$&$\scr{3133}^{(+)}$&&&&&\\
\hline
$\scr{22}$&$\scr{21}$&$\scr{2121}$&$\scr{2111}^{(+)}$&$\scr{1133}$&&&&\\
\hline
$\scr{23}$&$\scr{31}$&$\scr{3131}$&$\scr{3111}^{(+)}$&&&&&\\
\hline
$\scr{24}$&$\scr{42}$&$\scr{4242}$&$\scr{4222}^{(+)}$&$\scr{2233}$&&&&\\
\hline
$\scr{25}$&$\scr{52}$&$\scr{5252}$&$\scr{5222}^{(+)}$&&&&&\\
\hline
$\scr{26}$&$\scr{21}$&$\scr{2121}$&$\scr{2122}^{(+)}$&&&&&\\
\hline
$\scr{27}$&$\scr{97}$&$\scr{9797}$&$\scr{9777}^{(+)}$&$\scr{7733}$&&&&\\
\hline
$\scr{28}$&$\scr{92}$&$\scr{9297}$&$\scr{9777}^{(+)}$&$\scr{7733}$&&&&\\
\hline
$\scr{29}$&$\scr{127}$&$\scr{127127}$&$\scr{12777}^{(+)}$&&&&&\\
\hline
$\scr{30}$&$\scr{118}$&$\scr{118118}$&$\scr{11888}^{(+)}$&$\scr{8833}$&&&&\\
\hline
$\scr{31}$&$\scr{108}$&$\scr{108108}$&$\scr{10888}^{(+)}$&&&&&\\
\hline
$\scr{32}$&$\scr{103}$&$\scr{103108}$&$\scr{10888}^{(+)}$&&&&&\\
\hline
$\scr{33}$&$\scr{53}$&$\scr{5353}$&$\scr{5323}^{(+)}$&&&&&\\
\hline
$\scr{34}$&$\scr{21}$&$\scr{2121}$&$\scr{2123}^{(+)}$&&&&&\\
\hline
$\scr{35}$&$\scr{21}$&$\scr{2121}$&$\scr{2111}^{(+)}$&$\scr{1111}$&$\scr{1113}^{(+)}$&&&\\
\hline
$\scr{36}$&$\scr{31}$&$\scr{3131}$&$\scr{3111}^{(+)}$&&&&&\\
\hline
\hline
\caption{Pathways participating in the 2DIR vibrational spectrum with the laser field $a$ or $b$ acting as the first interaction.The symbol $(\pm)$ stand for the sign of the $\vec{k}$ component of the exciting laser fields.}
\end{longtable}}
\end{center}

Next, the second table is given by
\begin{center}
{\scriptsize
\begin{longtable}{c@{\phantom{---}}cccccccc}
\hline
\hline
Path &$\!\scr{\g\rho(t)}\!$&$\!\scr{\g{G}(t-\tau_3)}\!$&$\!\scr{\g{L}_{v[p]}(\tau_3)}\!$&$\!\scr{\g{G}(\tau_3-\tau_2)}\!$&$\!\scr{\g{L}_{v[a(b)]}(\tau_2)}\!$&$\!\scr{\g{G}((\tau_2-\tau_1))}\!$ &$\!\scr{\g{L}_{v[r]}(\tau_1)}\!$&$\!\scr{\g{\rho}(t_0)}\!$\\
\hline
\hline
$\scr{37}$&$\scr{42}$&$\scr{4242}$&$\scr{4222}^{(+)}$&$\scr{2222}$&$\scr{2221}^{(-)}$&$\scr{2121}$&$\scr{2111}^{(+)}$&$\scr{11}$\\
\hline
$\scr{38}$&$\scr{52}$&$\scr{5252}$&$\scr{5222}^{(+)}$&&&&&\\
\hline
$\scr{39}$&$\scr{21}$&$\scr{2121}$&$\scr{2122}^{(+)}$&&&&&\\
\hline
$\scr{40}$&$\scr{21}$&$\scr{2121}$&$\scr{2111}^{(+)}$&$\scr{1122}$&&&&\\
\hline
$\scr{41}$&$\scr{31}$&$\scr{3131}$&$\scr{3111}^{(+)}$&&&&&\\
\hline
$\scr{42}$&$\scr{53}$&$\scr{5353}$&$\scr{5333}^{(+)}$&$\scr{3322}$&&&&\\
\hline
$\scr{43}$&$\scr{63}$&$\scr{6363}$&$\scr{6333}^{(+)}$&&&&&\\
\hline
$\scr{44}$&$\scr{31}$&$\scr{3131}$&$\scr{3133}^{(+)}$&&&&&\\
\hline
$\scr{45}$&$\scr{97}$&$\scr{9797}$&$\scr{9777}^{(+)}$&$\scr{7722}$&&&&\\
\hline
$\scr{46}$&$\scr{92}$&$\scr{9297}$&$\scr{9777}^{(+)}$&$\scr{7722}$&&&&\\
\hline
$\scr{47}$&$\scr{127}$&$\scr{127127}$&$\scr{12777}^{(+)}$&&&&&\\
\hline
$\scr{48 }$&$\scr{118}$&$\scr{118118}$&$\scr{11888}^{(+)}$&$\scr{8822}$&&&&\\
\hline
$\scr{49}$&$\scr{108}$&$\scr{108108}$&$\scr{10888}^{(+)}$&&&&&\\
\hline
$\scr{50}$&$\scr{103}$&$\scr{103108}$&$\scr{10888}^{(+)}$&&&&&\\
\hline
$\scr{51}$&$\scr{53}$&$\scr{5353}$&$\scr{5323}^{(+)}$&&&&&\\
\hline
$\scr{52}$&$\scr{21}$&$\scr{2121}$&$\scr{2123}^{(+)}$&&&&&\\
\hline
$\scr{53}$&$\scr{21}$&$\scr{2121}$&$\scr{2111}^{(+)}$&$\scr{1111}$&$\scr{1121}^{(-)}$&&&\\
\hline
$\scr{54}$&$\scr{31}$&$\scr{3131}$&$\scr{3111}^{(+)}$&&&&&\\
\hline
$\scr{55}$&$\scr{53}$&$\scr{5353}$&$\scr{5333}^{(+)}$&$\scr{3333}$&$\scr{3331}^{(-)}$&$\scr{3131}$&$\scr{3111}^{(+)}$&\\
\hline
$\scr{56}$&$\scr{63}$&$\scr{6363}$&$\scr{6333}^{(+)}$&&&&&\\
\hline
$\scr{57}$&$\scr{31}$&$\scr{3131}$&$\scr{3133}^{(+)}$&&&&&\\
\hline
$\scr{58}$&$\scr{21}$&$\scr{2121}$&$\scr{2111}^{(+)}$&$\scr{1133}$&&&&\\
\hline
$\scr{59}$&$\scr{31}$&$\scr{3131}$&$\scr{3111}^{(+)}$&&&&&\\
\hline
$\scr{60}$&$\scr{42}$&$\scr{4242}$&$\scr{4222}^{(+)}$&$\scr{2233}$&&&&\\
\hline
$\scr{61}$&$\scr{52}$&$\scr{5252}$&$\scr{5222}^{(+)}$&&&&&\\
\hline
$\scr{62}$&$\scr{21}$&$\scr{2121}$&$\scr{2122}^{(+)}$&&&&&\\
\hline
$\scr{63}$&$\scr{97}$&$\scr{9797}$&$\scr{9777}^{(+)}$&$\scr{7733}$&&&&\\
\hline
$\scr{64}$&$\scr{92}$&$\scr{9297}$&$\scr{9777}^{(+)}$&$\scr{7733}$&&&&\\
\hline
$\scr{65}$&$\scr{127}$&$\scr{127127}$&$\scr{12777}^{(+)}$&&&&&\\
\hline
$\scr{66}$&$\scr{118}$&$\scr{118118}$&$\scr{11888}^{(+)}$&$\scr{8833}$&&&&\\
\hline
$\scr{67}$&$\scr{108}$&$\scr{108108}$&$\scr{10888}^{(+)}$&&&&&\\
\hline
$\scr{68}$&$\scr{103}$&$\scr{103108}$&$\scr{10888}^{(+)}$&&&&&\\
\hline
$\scr{69}$&$\scr{52}$&$\scr{5252}$&$\scr{5232}^{(+)}$&$\scr{3232}$&$\scr{3231}^{(-)}$&&&\\
\hline
$\scr{70}$&$\scr{31}$&$\scr{3131}$&$\scr{3132}^{(+)}$&&&&&\\
\hline
$\scr{71}$&$\scr{21}$&$\scr{2121}$&$\scr{2111}^{(+)}$&$\scr{1111}$&$\scr{1131}^{(-)}$&&&\\
\hline
$\scr{72}$&$\scr{31}$&$\scr{3131}$&$\scr{3111}^{(+)}$&&&&&\\
\hline
\hline
\caption{Pathways participating in the 2D vibrational spectrum with the laser field $a$ or $b$ acting as the second interaction.}
\end{longtable}}
\end{center}
Finally, the last table takes the form
\begin{center}
{\scriptsize
\begin{longtable}{c@{\phantom{---}}cccccccc}
\hline
\hline
Path &$\!\scr{\g\rho(t)}\!$&$\!\scr{\g{G}(t-\tau_3)}\!$&$\!\scr{\g{L}_{v[a(b)]}(\tau_3)}\!$&$\!\scr{\g{G}(\tau_3-\tau_2)}\!$&$\!\scr{\g{L}_{v[q]}(\tau_2)}\!$&$\!\scr{\g{G}((\tau_2-\tau_1))}\!$ &$\!\scr{\g{L}_{v[r]}(\tau_1)}\!$&$\!\scr{\g{\rho}(t_0)}\!$\\
\hline
\hline
$\scr{73}$&$\scr{21}$&$\scr{2121}$&$\scr{2141}^{(-)}$&$\scr{4141}$&$\scr{4121}^{(+)}$&$\scr{2121}$&$\scr{2111}^{(+)}$&$\scr{11}$\\
\hline
$\scr{74}$&$\scr{42}$&$\scr{4242}$&$\scr{4241}^{(-)}$&&&&&\\
\hline
$\scr{75}$&$\scr{21}$&$\scr{2121}$&$\scr{2151}^{(-)}$&$\scr{5151}$&$\scr{5121}^{(+)}$&&&\\
\hline
$\scr{76}$&$\scr{31}$&$\scr{3131}$&$\scr{3151}^{(-)}$&&&&&\\
\hline
$\scr{77}$&$\scr{52}$&$\scr{5252}$&$\scr{5251}^{(-)}$&&&&&\\
\hline
$\scr{78}$&$\scr{53}$&$\scr{5353}$&$\scr{5351}^{(-)}$&&&&&\\
\hline
$\scr{79}$&$\scr{31}$&$\scr{3131}$&$\scr{3161}^{(-)}$&$\scr{6161}$&$\scr{6131}^{(+)}$&$\scr{3131}$&$\scr{3111}^{(+)}$&\\
\hline
$\scr{78}$&$\scr{63}$&$\scr{6363}$&$\scr{6361}^{(-)}$&&&&&\\
\hline
$\scr{79}$&$\scr{21}$&$\scr{2121}$&$\scr{2151}^{(-)}$&$\scr{5151}$&$\scr{5131}^{(+)}$&&&\\
\hline
$\scr{80}$&$\scr{31}$&$\scr{3131}$&$\scr{3151}^{(-)}$&&&&&\\
\hline
$\scr{81}$&$\scr{52}$&$\scr{5252}$&$\scr{5251}^{(-)}$&&&&&\\
\hline
$\scr{82}$&$\scr{53}$&$\scr{5353}$&$\scr{5351}^{(-)}$&&&&&\\
\hline
\hline
\caption{Pathways participating in the 2D vibrational spectrum with the laser field $a$ or $b$ acting as the third interaction.}
\end{longtable}}
\end{center}

\section{Time-orderd integrals}

First, we introduce various constants useful for simplify the final expressions
\begin{eqnarray}
&&\mbox{$\scr  K_T=\sqrt{\gamma_a\gamma_b\gamma_c}\e^{i\omega_aT-i\omega_bT} $} \hskip 1.45truecm   \hbox{$\scr  TF_{\omega_{lo}}(\omega_t)=\frac{\e^{i\omega_tT_{lo}}}{i\omega_t-i\omega_{lo}+\gamma_{lo}}-\frac{\e^{i\omega_tT_{lo}}}{i\omega_t-i\omega_{lo}-\gamma_{lo}} $}  \nonumber\\
&&\hbox{$\scr  C_0^+=\frac{1}{i\omega_d-i\omega_a+\gamma_a}-\frac{1}{i\omega_d-i\omega_a-\gamma_a}    \qquad  C_0^-=\frac{1}{i\omega_d-i\omega_b-\gamma_b}-\frac{1}{i\omega_d-i\omega_b+\gamma_b} $}  .
\label{B.1}
\end{eqnarray}
Then, for the laser field interaction ordering $c\leftarrow b\leftarrow a$ and a positive delay time $T_b-T_a>0$, there are
\begin{eqnarray}
&&\mbox{$\scr C_{1,cba}^+[n,\alpha]=\;\;\;\frac{C_0^+\e^{-C[n,\alpha,a]T}}{C[n,\alpha,a]+i\omega_d-i\omega_a}-\frac{\e^{-C[n,\alpha,a]T}}{(C[n,\alpha,a]-\gamma_a)(i\omega_d-i\omega_a+\gamma_a)}+\frac{\e^{-C[n,\alpha,a]T}}{(C[n,\alpha,a]+\gamma_a)(i\omega_d-i\omega_a-\gamma_a)}   $}  \nonumber\\
&&\mbox{$\scr C_{2,cba}^+[n,\alpha]=-\frac{\e^{-B[n,\alpha,a,b]T}-C[n,\alpha,a]T}{(B[n,\alpha,a,b]-\gamma_b+C[n,\alpha,a]-\gamma_a)(C[n,\alpha,a]-\gamma_a)(i\omega_d-i\omega_a+\gamma_a)}-\frac{C_{1,cba}^+[n,\alpha]\e^{-B[n,\alpha,a,b]T}}{B[n,\alpha,a,b]-\gamma_b}  $}        \nonumber\\
&&\mbox{$\scr\hskip 1.8truecm +\frac{C_0^+\e^{-B[n,\alpha,a,b]T-C[n,\alpha,a]T}}{(B[n,\alpha,a,b]+\gamma_b+C[n,\alpha,a]+i\omega_d-i\omega_a)(C[n,\alpha,a]+i\omega_d-i\omega_a)}+\frac{\e^{-B[n,\alpha,a,b]T-C[n,\alpha,a]T}}{(B[n,\alpha,a,b]+\gamma_b+C[n,\alpha,a]+\gamma_a)(C[n,\alpha,a]+\gamma_a)(i\omega_d-i\omega_a-\gamma_a)} $}     \nonumber\\
&&\mbox{$\scr C_{3,cba,1}^+[n,\alpha]=-\frac{\e^{-\gamma_aT-\gamma_bT}}{(A[n,\alpha,a,b,c]-\gamma_c+B[n,\alpha,a,b]-\gamma_b+C[n,\alpha,a]-\gamma_a)(B[n,\alpha,a,b]-\gamma_b+C[n,\alpha,a]-\gamma_a)(C[n,\alpha,a]-\gamma_a)(i\omega_d-i\omega_a+\gamma_a)}    $}  \nonumber\\
&&\mbox{$\scr\hskip 1.8truecm +\frac{\e^{-\gamma_aT-\gamma_bT}}{(A[n,\alpha,a,b,c]+\gamma_c+B[n,\alpha,a,b]-\gamma_b+C[n,\alpha,a]-\gamma_a)(B[n,\alpha,a,b]-\gamma_b+C[n,\alpha,a]-\gamma_a)(C[n,\alpha,a]-\gamma_a)(i\omega_d-i\omega_a+\gamma_a)}   $}  \nonumber\\
&&\mbox{$\scr\hskip 1.8truecm -\frac{C_{1,cba}^{+}[n,\alpha]\e^{-\gamma_bT}}{(A[n,\alpha,a,b,c]-\gamma_c+B[n,\alpha,a,b]-\gamma_b)(B[n,\alpha,a,b]-\gamma_b)} +\frac{C_{1,cba}^{+}[n,\alpha]\e^{-\gamma_bT}}{(A[n,\alpha,a,b,c]+\gamma_c+B[n,\alpha,a,b]-\gamma_b)(B[n,\alpha,a,b]-\gamma_b)}    $}  \nonumber\\
&&\mbox{$\scr\hskip 1.8truecm  -\frac{C_{2,cba}^+[n,\alpha]}{A[n,\alpha,a,b,c]-\gamma_c}+\frac{C_{2,cba}^+[n,\alpha]}{A[n,\alpha,a,b,c]+\gamma_c} $}  \nonumber\\
&&\mbox{$\scr C_{3,cba,2}^+[n,\alpha]=-\frac{\e^{-A[n,\alpha,a,b,c]T-B[n,\alpha,a,b]T-C[n,\alpha,a]T-\gamma_cT}}{(A[n,\alpha,a,b,c]+\gamma_c+B[n,\alpha,a,b]-\gamma_b+C[n,\alpha,a]-\gamma_a)(B[n,\alpha,a,b]-\gamma_b+C[n,\alpha,a]-\gamma_a)(C[n,\alpha,a]-\gamma_a)(i\omega_d-i\omega_a+\gamma_a)} $}  \nonumber\\
&&\mbox{$\scr\hskip 1.8truecm +\frac{C_0^+\e^{-A[n,\alpha,a,b,c]T-B[n,\alpha,a,b]T-C[n,\alpha,a]T-\gamma_cT}}{(A[n,\alpha,a,b,c]+\gamma_c+B[n,\alpha,a,b]+\gamma_b+C[n,\alpha,a]+i\omega_d-i\omega_a)(B[n,\alpha,a,b]+\gamma_b+C[n,\alpha,a]+i\omega_d-i\omega_a)(C[n,\alpha,a]+i\omega_d-i\omega_a)}   $}  \nonumber\\
&&\mbox{$\scr\hskip 1.8truecm +\frac{\e^{-A[n,\alpha,a,b,c]T-B[n,\alpha,a,b]T-C[n,\alpha,a]T-\gamma_cT}}{(A[n,\alpha,a,b,c]+\gamma_c+B[n,\alpha,a,b]+\gamma_b+C[n,\alpha,a]+\gamma_a)(B[n,\alpha,a,b]+\gamma_b+C[n,\alpha,a]+\gamma_a)(C[n,\alpha,a]+\gamma_a)(i\omega_d-i\omega_a-\gamma_a)} $}  \nonumber\\
&&\mbox{$\scr\hskip 1.8truecm -\frac{C_{1,cba}^+[n,\alpha]\e^{-A[n,\alpha,a,b,c]T-B[n,\alpha,a,b]T-\gamma_cT}}{(A[n,\alpha,a,b,c]+\gamma_c+B[n,\alpha,a,b]-\gamma_b)(B[n,\alpha,a,b]-\gamma_b)}-\frac{C_{2,cba}^+[n,\alpha]\e^{-A[n,\alpha,a,b,c]T-\gamma_cT}}{A[n,\alpha,a,b,c]+\gamma_c}   $}  
\label{B.2}
\end{eqnarray}
and the first contribution for positive delay time $T_b-T_a$ and field ordering $c\leftarrow b\leftarrow a$ is given by
\begin{eqnarray}
&&\mbox{$\scr I_{cba}^+[n,\omega_d,\omega_t]=-\sum_{\alpha}TF_{\omega_{lo}}(\omega_t)Q[n,\alpha,a,b,c]K_T $}  \nonumber\\ 
&&\mbox{$\scr\hskip 1.4truecm\times\Biggl\lbrack\;\;\; 
\frac{\e^{-\gamma_aT-\gamma_bT}}{(A[n,\alpha,a,b,c]-\gamma_c+B[n,\alpha,a,b]-\gamma_b+C[n,\alpha,a]-\gamma_a)(B[n,\alpha,a,b]-\gamma_b+C[n,\alpha,a]-\gamma_a)(C[n,\alpha,a]-\gamma_a)}$}  \nonumber\\
&&\mbox{$\scr\hskip 2.4truecm\times
\frac{1}{(i\omega_d-i\omega_a+\gamma_a)(K[n,\alpha,a,b,c]+i\omega_t+A[n,\alpha,a,b,c]-\gamma_c+B[n,\alpha,a,b]-\gamma_b+C[n,\alpha,a]-\gamma_a)} $}  \nonumber\\
&&\mbox{$\scr\hskip 1.8truecm +\frac{C_{1,cba}^+[n,\alpha]\e^{-\gamma_bT}}{(A[n,\alpha,a,b,c]-\gamma_c+B[n,\alpha,a,b]-\gamma_b)(B[n,\alpha,a,b]-\gamma_b)(K[n,\alpha,a,b,c]+i\omega_t+A[n,\alpha,a,b,c] -\gamma_c+B[n,\alpha,a,b]-\gamma_b)}   $}  \nonumber\\
&&\mbox{$\scr\hskip 1.8truecm +\frac{C_{2,cba}^+[n,\alpha]}{(A[n,\alpha,a,b,c]-\gamma_c)(K[n,\alpha,a,b,c]+i\omega_t+A[n,\alpha,a,b,c]-\gamma_c)} +\frac{C_{3,cba,1}^+[n,\alpha]}{K[n,\alpha,a,b,c]+i\omega_t} +\frac{C_{3,cba,2}^+[n,\alpha]}{K[n,\alpha,a,b,c]+i\omega_t}$}  \Biggr\rbrack  .
\label{B.3}
\end{eqnarray}
With the same ordering of the laser pulses, say $c\leftarrow b\leftarrow a$ and a negative delay time $T_b-T_a<0$, it is found that
\begin{eqnarray}
&&\mbox{$\scr C_{1,cba}^-[n,\alpha]=\;\;\;\frac{\e^{-C[n,\alpha,a]T}}{C[n,\alpha,a]-\gamma_a}-\frac{\e^{-C[n,\alpha,a]T}}{(C[n,\alpha,a]+\gamma_a)}       $}  \nonumber\\
&&\mbox{$\scr C_{2,cba}^-[n,\alpha]=\;\;\;\frac{\e^{-B[n,\alpha,a,b]T-C[n,\alpha,a]T}}{(B[n,\alpha,a,b]-\gamma_a+C[n,\alpha,a]-\gamma_b)(C[n,\alpha,a]-\gamma_a)(i\omega_d-i\omega_b-\gamma_b)}-\frac{C_{1,cba}^-[n,\alpha]\e^{-B[n,\alpha,a,b]T}}{(B[n,\alpha,a,b]-\gamma_b)(i\omega_d-i\omega_b-\gamma_b)}  $}        \nonumber\\
&&\mbox{$\scr\hskip 1.6truecm +\frac{C_0^-\e^{-B[n,\alpha,a,b]T-C[n,\alpha,a]T}}{(-B[n,\alpha,a,b]-C[n,\alpha,a]-\gamma_a +i\omega_d-i\omega_b)(C[n,\alpha,a] +\gamma_a)}-\frac{\e^{-B[n,\alpha,a,b]T-C[n,\alpha,a]T}}{(B[n,\alpha,a,b]+C[n,\alpha,a]+\gamma_a+\gamma_b)(C[n,\alpha,a]+\gamma_a)(i\omega_d-i\omega_b+\gamma_b)} $}   \nonumber\\
&&\mbox{$\scr C_{3,cba,1}^-[n,\alpha]=\;\;\;\frac{\e^{-\gamma_aT-\gamma_bT}}{(A[n,\alpha,a,b,c]-\gamma_c+B[n,\alpha,a,b]-\gamma_a+C[n,\alpha,a]-\gamma_b)(B[n,\alpha,a,b]-\gamma_a+C[n,\alpha,a]-\gamma_b)(C[n,\alpha,a]-\gamma_a)(i\omega_d-i\omega_b-\gamma_b)}    $}  \nonumber\\
&&\mbox{$\scr\hskip 1.9truecm -\frac{\e^{-\gamma_aT-\gamma_bT}}{(A[n,\alpha,a,b,c]+\gamma_c+B[n,\alpha,a,b]-\gamma_a+C[n,\alpha,a]-\gamma_b)(B[n,\alpha,a,b]-\gamma_a+C[n,\alpha,a]-\gamma_b)(C[n,\alpha,a]-\gamma_a)(i\omega_d-i\omega_b-\gamma_b)}  $}  \nonumber\\
&&\mbox{$\scr\hskip 1.9truecm  -\frac{C_{2,cba}^-[n,\alpha]}{A[n,\alpha,a,b,c]-\gamma_c} +\frac{C_{2,cba}^-[n,\alpha]}{A[n,\alpha,a,b,c]+\gamma_c}  -\frac{C_{1,cba}^-[n,\alpha]\e^{-\gamma_bT}}{(A[n,\alpha,a,b,c]-\gamma_c+B[n,\alpha,a,b]-\gamma_b)(B[n,\alpha,a,b]-\gamma_b)(i\omega_d-i\omega_b+\gamma_b)}$}  \nonumber\\
&&\mbox{$\scr\hskip 1.9truecm 
+\frac{C_{1,cba}^-[n,\alpha]\e^{-\gamma_bT}}{(A[n,\alpha,a,b,c]+\gamma_c+B[n,\alpha,a,b]-\gamma_b)(B[n,\alpha,a,b]-\gamma_b)(i\omega_d-i\omega_b-\gamma_b)}$}  \nonumber\\
&&\mbox{$\scr C_{3,cba,2}^-[n,\alpha]=\;\;\;\frac{\e^{-A[n,\alpha,a,b,c]T-B[n,\alpha,a,b]T-C[n,\alpha,a]T-\gamma_cT}}{(A[n,\alpha,a,b,c]+\gamma_c+B[n,\alpha,a,b]-\gamma_a+C[n,\alpha,a]-\gamma_b)(B[n,\alpha,a,b]-\gamma_a+C[n,\alpha,a]-\gamma_b)(C[n,\alpha,a]-\gamma_a)(i\omega_d-i\omega_b-\gamma_b)} $}  \nonumber\\
&&\mbox{$\scr\hskip 1.9truecm -\frac{C_0^-\e^{-A[n,\alpha,a,b,c]T-B[n,\alpha,a,b]T-C[n,\alpha,a]T-\gamma_cT}}{(-A[n,\alpha,a,b,c]-\gamma_c-B[n,\alpha,a,b]-C[n,\alpha,a]-\gamma_a+i\omega_d-i\omega_b)(-B[n,\alpha,a,b]-C[n,\alpha,a]-\gamma_a+i\omega_d-i\omega_b)(C[n,\alpha,a]T+\gamma_a)}   $}  \nonumber\\
&&\mbox{$\scr\hskip 1.9truecm -\frac{\e^{-A[n,\alpha,a,b,c]T-B[n,\alpha,a,b]T-C[n,\alpha,a]T-\gamma_cT}}{(A[n,\alpha,a,b,c]+\gamma_c+B[n,\alpha,a,b]+C[n,\alpha,a]+\gamma_a+\gamma_b)(B[n,\alpha,a,b]+C[n,\alpha,a]+\gamma_a+\gamma_b)(C[n,\alpha,a]+\gamma_a)(i\omega_d-i\omega_b+\gamma_b)} $}  \nonumber\\
&&\mbox{$\scr\hskip 1.9truecm -\frac{C_{1,cba}^-[n,\alpha]\e^{-A[n,\alpha,a,b,c]T-B[n,\alpha,a,b]T-\gamma_cT}}{(A[n,\alpha,a,b,c]+\gamma_c+B[n,\alpha,a,b]-\gamma_b)(B[n,\alpha,a,b]-\gamma_b)(i\omega_d-i\omega_b-\gamma_b)}-\frac{C_{2,cba}^-[n,\alpha]\e^{-A[n,\alpha,a,b,c]T-\gamma_cT}}{A[n,\alpha,a,b,c]+\gamma_c}   $}  
\label{B.4}
\end{eqnarray}
and the second contribution for negative delay time $T_b-T_a<0$ is
\begin{eqnarray}
&&\mbox{$\scr I_{cba}^-[n,\omega_d,\omega_t]=-\sum_{\alpha}TF_{\omega_{lo}}(\omega_t)Q[n,\alpha,a,b,c]K_T $}  \nonumber\\ 
&&\mbox{$\scr\hskip 0.8truecm\times\Biggl\lbrack\;\;\;
\frac{\e^{-\gamma_aT-\gamma_bT}}{(A[n,\alpha,a,b,c]-\gamma_c+B[n,\alpha,a,b]-\gamma_a+C[n,\alpha,a]-\gamma_b)(B[n,\alpha,a,b]-\gamma_a+C[n,\alpha,a]-\gamma_b)(C[n,\alpha,a]-\gamma_a)}  $}  \nonumber\\
&&\mbox{$\scr\hskip 1.8truecm\times
\frac{1}{(i\omega_d-i\omega_b-\gamma_b)(K[n,\alpha,a,b,c]+i\omega_t+A[n,\alpha,a,b,c]-\gamma_c+B[n,\alpha,a,b]+C[n,\alpha,a]-\gamma_a-\gamma_b)} $}  \nonumber\\
&&\mbox{$\scr\hskip 1.3truecm +\frac{C_{1,cba}^-[n,\alpha]\e^{-\gamma_bT}}{(A[n,\alpha,a,b,c]-\gamma_c+B[n,\alpha,a,b]-\gamma_b)(B[n,\alpha,a,b]-\gamma_b)(i\omega_d-i\omega_b-\gamma_b)                 (K[n,\alpha,a,b,c]+i\omega_t+A[n,\alpha,a,b,c]-\gamma_c+B[n,\alpha,a,b]-\gamma_b)}   $}  \nonumber\\
&&\mbox{$\scr\hskip 1.3truecm +\frac{C_{2,cba}^-[n,\alpha]}{(A[n,\alpha,a,b,c]-\gamma_c)(K[n,\alpha,a,b,c]+i\omega_t+A[n,\alpha,a,b,c]-\gamma_c)} +\frac{C_{3,cba,1}^-[n,\alpha]}{K[n,\alpha,a,b,c]+i\omega_t} +\frac{C_{3,cba,2}^-[n,\alpha]}{K[n,\alpha,a,b,c]+i\omega_t}   $}  \Biggr\rbrack
\label{B.5}
\end{eqnarray}
and the contribution to the 2D-IR spectrum with field ordering $c\leftarrow b\leftarrow a$ is given by 
\begin{eqnarray}
I_{cba}[n,\omega_d,\omega_t]=H(T_b-T_a)I_{cba}^+[n,\omega_d,\omega_t]+H(T_a-T_b)I_{cba}^-[n,\omega_d,\omega_t]
\label{B.6}
\end{eqnarray}
where $H(t)$ stands for the Heaviside function. The contributions for any other field ordering can be evaluated as well.

\end{document}